\documentclass[aps,prd,notitlepage,showpacs,nofootinbib,superscriptaddress]{revtex4-1}
\pdfoutput=1

\usepackage{graphicx}
\usepackage[utf8]{inputenc}

\usepackage{dsfont}
\usepackage{amsmath}
\usepackage{amssymb}
\usepackage{amsfonts}
\usepackage{float}
\usepackage{comment}
\usepackage{slashed}
\usepackage{bbm}
\usepackage{bbold}

\usepackage{booktabs}

\usepackage{mathtools,braket,blkarray}

\usepackage{amsmath}
\usepackage{amssymb}

\usepackage{soul}

\usepackage[usenames,dvipsnames]{color}
\usepackage[colorinlistoftodos]{todonotes}
\usepackage[colorlinks=true,citecolor=darkred,urlcolor=darkred, pdfborder={0 0 0}]{hyperref}

\usepackage{multirow}
\definecolor{darkred}{rgb}{0.6,0,0}
\definecolor{darkgreen}{rgb}{0.0, 0.4, 0.0}
\usepackage[colorinlistoftodos]{todonotes}

\definecolor{linkcolor}{rgb}{0,0,0.5}

\newcommand {\ignore}[1]{}

\def \znbb {$\rm 0\nu\beta\beta$ }

\def\gsim{\raise0.3ex\hbox{$\;>$\kern-0.75em\raise-1.1ex\hbox{$\sim\;$}}}
\def\lsim{\raise0.3ex\hbox{$\;<$\kern-0.75em\raise-1.1ex\hbox{$\sim\;$}}}

\def\SM{$\mathrm{SU(3)_c \otimes SU(2)_L \otimes U(1)_Y}$ }
\newcommand{\sm}{{Standard Model }}

\usepackage{braket}
\usepackage{cancel}

\usepackage{soul}

\definecolor{mightnightblue}{RGB}{25,25,112}

\definecolor{brown}{rgb}{0.59, 0.29, 0.0}

\def\SM{$\mathrm{SU(3)_c \otimes SU(2)_L \otimes U(1)_Y}$ }
\def\21{$\mathrm{SU(2)_L \otimes U(1)_Y}$}

\def\sm{standard model }

\newcommand{\AddrAHEP}{%
  AHEP Group, Institut de F\'{i}sica Corpuscular --
  C.S.I.C./Universitat de Val\`{e}ncia, Parc Cient\'ific de Paterna.\\
 C/ Catedr\'atico Jos\'e Beltr\'an, 2 E-46980 Paterna (Valencia) - SPAIN}

\begin{document}

\begin{flushright}
{\tt USTC-ICTS/PCFT-20-27}
\end{flushright}

\bibliographystyle{unsrt}   

\title{\boldmath \color{BrickRed} Trimaximal neutrino mixing from scotogenic $A_4$ family symmetry}

\author{Gui-Jun Ding}\email{dinggj@ustc.edu.cn}
\affiliation{
Peng Huanwu Center for Fundamental Theory, Hefei, Anhui 230026, China}
\affiliation{
Interdisciplinary Center for Theoretical Study and Department of Modern Physics, \\
University of Science and Technology of China, Hefei, Anhui 230026, China}
\author{Jun-Nan Lu}\email{JunNan.Lu@ific.uv.es}
\affiliation{
Interdisciplinary Center for Theoretical Study and Department of Modern Physics, \\
University of Science and Technology of China, Hefei, Anhui 230026, China}
\affiliation{\AddrAHEP}
\author{Jos\'{e} W. F. Valle}\email{valle@ific.uv.es}
\affiliation{\AddrAHEP}

\begin{abstract}
  \vspace{1cm}

 We propose a flavour theory of leptons implementing an $A_4$ family symmetry.
Our scheme provides a simple way to derive trimaximal neutrino mixing from first principles, leading to simple and testable predictions for neutrino mixing and CP violation.
Dark matter mediates neutrino mass generation, as in the simplest scotogenic model.

\end{abstract}

\maketitle


\section{Introduction}
\label{sec:introduction}


The discovery of neutrino oscillations~\cite{McDonald:2016ixn,Kajita:2016cak,deSalas:2020pgw} constitutes a milestone in particle physics, yet we are still far from understanding the pattern of lepton mixing parameters.
Indeed, the flavor problem of particle physics has, in the lepton sector, its most challenging expression.
Likewise, the basic understanding and interpretaion of cosmological dark matter remains a challenge~\cite{Bertone:2004pz}. Many educated guesses have been proposed as to what the pattern of lepton mixing should look like.
Such phenomenological patterns of neutrino mixing include, for example, Tri-bimaximal (TBM)~\cite{Harrison:2002er,Harrison:2002kp}, Trimaximal (TM1/TM2)~\cite{Albright:2008rp,Albright:2010ap,He:2011gb} and bi-large mixing patterns~\cite{Boucenna:2012xb,Roy:2014nua,Chen:2019egu,Ding:2019vvi}.
Some of these have been generalized~\cite{Datta:2005ci}, especially after~\cite{Chen:2018eou,Chen:2019fgb} the non-zero value of the mixing angle $\theta_{13}$ was established by reactor experiments Daya Bay~\cite{An:2016ses} and RENO~\cite{Pac:2018scx}.

Extracting the symmetries behind the observed pattern of neutrino oscillations seems to be a promising way to make progress~\cite{Ishimori:2010au}.
Such symmetry programme may be pursued in a model-independent way, for example exploiting generalized CP symmetries~\cite{Chen:2015siy,Chen:2016ica,Chen:2018lsv,Chen:2018zbq,Chen:2018zbq},
or attempts to implement full-fledged family symmetries from first principles~\cite{Babu:2002dz,Altarelli:2005yx,Altarelli:2005yp,Chen:2015jta,deAnda:2019jxw,deAnda:2020pti,Perez:2019aqq,Chen:2020udk}.
This way one can obtain neutrino mixing predictions within fundamental theories of neutrino mass.\\

The aim of this letter is to pursue the symmetry approach to account for the oscillation results within a scenario in which the dark matter problem is also addressed.
We propose a neutrino mass theory implementing an $A_4$ family symmetry in which the dark matter particle is identified with one of the mediators of neutrino mass generation.
Such ``scotogenic'' paradigm~\cite{Ma:2006km} has been generalized in many ways~\cite{Kang:2019sab,Leite:2019grf,CarcamoHernandez:2020ehn,Leite:2020bnb,Leite:2020wjl}.
Here we focus on an \SM extension of the simplest scotogenic model, in which the dark fermions include both $\mathrm{SU(2)_L}$ iso-singlets~\cite{Ma:2006km} as well as iso-triplets~\cite{Ma:2008cu}.
Such ``cloned'' model was suggested in~\cite{Hirsch:2013ola} and has a richer phenomenology, studied in a number of papers~\cite{Hirsch:2013ola,Merle:2016scw,Diaz:2016udz,Choubey:2017yyn,Restrepo:2019ilz,Avila:2019hhv}.
Our proposed flavour extension of this scenario enables us to derive, from first principles, the TM2 neutrino mixing ansatz~\cite{Albright:2008rp,Albright:2010ap,He:2011gb}.
This results in two very simple predictions for neutrino mixing and CP violation parameters, that will be tested at upcoming neutrino experiments.

\section{A simple scotogenic model with $A_4$ family symmetry}
\label{sec:model}

Here we extend the singlet-triplet scotogenic model of~\cite{Hirsch:2013ola} by combining with the $A_4$ flavor symmetry in order to understand the observed flavor structure in the lepton sector.
We recall that $A_4$ is the even permutation group of four objects, and it can be generated by two generators $S$ and $T$ obeying the relations
\begin{equation}
S^2=T^3=(ST)^3=1\,.
\end{equation}
$A_4$ has three irreducible one-dimensional $\mathbf{1}$, $\mathbf{1'}$ and $\mathbf{1''}$ and a three-dimensional representation $\mathbf{3}$. We formulate the model in the basis where both generators $S$ and $T$ in the triplet representation $\mathbf{3}$ are represented by real matrices~\cite{Ma:2001dn},
\begin{equation}
S=\begin{pmatrix}
1&0&0\\
0&-1&0\\
0&0&-1
\end{pmatrix},~~~T=\begin{pmatrix}
0&1&0\\
0&0&1\\
1&0&0
\end{pmatrix}\,.
\end{equation}
The most relevant contraction rule is $\mathbf{3}\otimes\mathbf{3}=\mathbf{1}\oplus\mathbf{1}'\oplus\mathbf{1}''\oplus\mathbf{3}_{S}\oplus\mathbf{3}_{A}$, where $\mathbf{3}_{S}$ and $\mathbf{3}_{A}$ denote the symmetric and the antisymmetric triplet combinations respectively. For two triplets $\alpha= (\alpha_1, \alpha_2, \alpha_3)$ and
$\beta=(b_1,b_2,b_3)$, we have
\begin{eqnarray}
\nonumber && (\alpha\otimes\beta)_{\mathbf{1}}=\alpha_1\beta_1+\alpha_2\beta_2+\alpha_3\beta_3\,,\\
\nonumber&& (\alpha\otimes\beta)_{\mathbf{1}'}=\alpha_1\beta_1+\omega^2\alpha_2\beta_2+\omega\alpha_3\beta_3\,,\\
\nonumber&&(\alpha\otimes\beta)_{\mathbf{1}''}=\alpha_1\beta_1+\omega\alpha_2\beta_2+\omega^2\alpha_3\beta_3\,,\\
\nonumber&&(\alpha\otimes\beta)_{\mathbf{3}_{S}}=\left(
\begin{array}{c}\alpha_2\beta_3+\alpha_3\beta_2\\ \alpha_3\beta_1+\alpha_1\beta_3\\ \alpha_1\beta_2+\alpha_2\beta_1\end{array}\right)\,,\\
&&(\alpha\otimes\beta)_{\mathbf{3}_{A}}=\left(
\begin{array}{c}\alpha_2\beta_3-\alpha_3\beta_2\\ \alpha_3\beta_1-\alpha_1\beta_3\\ \alpha_1\beta_2-\alpha_2\beta_1\end{array}\right)\,.
\end{eqnarray}
\begin{table}[t!]
\centering
\begin{tabular}{|c|c|c|c|c|c|c|c| }
\hline
& \multicolumn{2}{ c |  }{~~Standard Model~~ } &  \multicolumn{2}{ c | }{ ~new fermions~ }  & \multicolumn{3}{c|}{~new scalars~}  \\
\hline
 &~   $L$   ~& $e_R$, $\mu_R$, $\tau_R$     & ~ $\Sigma$ ~&    $F$   &~  $\phi$  ~&~ $\eta$ ~&~ $\Omega$~ \\
\hline
Generations &   3 &  3 &  3 &  3 &  3 &  1 & 3\\
\hline\hline
$\rm SU(3)_C$  &   1    & 1       &   1     &    1    &  1  &    1    &   1    \\
$\rm SU(2)_L$  &  2    &  1        &    3     &   1  &    2   &    2    &   3     \\
$\rm U(1)_Y$     &   $-1$    & $-2$         &   0     &   0   &   1  & 1    &    0     \\  \hline
$\mathbb{Z} _{2}$      & $1$     & $1$    & $-1$    & $-1$  & $1$  & $-1$   &  $1$    \\
  \hline
$A_4$  &  $\mathbf{3}$  &  $\mathbf{1}, \mathbf{1'}, \mathbf{1''} $  & $\mathbf{3}$  &  $\mathbf{3}$  &  $\mathbf{3}$  &  $\mathbf{1}$  &  $\mathbf{3}$  \\
\hline \hline
\end{tabular}
\caption{Particle content and quantum numbers of our scotogenic model. }
\label{tab:fields_charges_first_version}
\end{table}

  In the original triplet-singlet scotogenic dark matter model~\cite{Hirsch:2013ola} there was only one copy of the new fermions $\Sigma$, $F$, as well as the new scalar fields $\eta$ and $\Omega$
  beyond the \sm Higgs doublet $\phi$.
  As a result, both charged lepton and neutrino mass matrices are structureless, with no prediction for lepton mixing~\cite{Hirsch:2013ola}.
  Here we assume three copies of $\Sigma$, $F$ and $\phi$, $\Omega$,  all transforming as the $A_4$ family group triplet $\mathbf{3}$, like the left-handed leptons.
The full particle content of the model is given in Table~\ref{tab:fields_charges_first_version}, with the corresponding assignments under the different symmetry groups.
Notice that the $\mathbb{Z} _{2}$ parity is imposed to ensure the stability of dark matter candidate.
Taking into account the new fields and symmetries of the model, the relevant terms of the Lagrangian involving fermion fields read
\begin{eqnarray}
  \nonumber
  \mathcal{L} & \supset & -y_e (\overline{L} \phi )_{\mathbf{1}} e_R-y_{\mu} (\overline{L} \phi )_{\mathbf{1''}} \mu_R-y_{\tau} (\overline{L} \phi )_{\mathbf{1'}} \tau_R-Y_{F}\left(\overline{L}F\right)_{\mathbf{1}}\tilde{\eta} - Y_{\Sigma}(\overline{L}\tilde{\Sigma}^{c})_{\mathbf{1}}\tilde{\eta}\\
  \label{eq:Lagrangian-fermion}  &~& -Y_{\Omega,1}\left( {\rm Tr} \left[\left(\overline{\Sigma}\Omega\right)_{\mathbf{3}_{S}} \right]F\right)_{\mathbf{1}}-Y_{\Omega,2}\left( {\rm Tr} \left[\left(\overline{\Sigma}\Omega\right)_{\mathbf{3}_{A}} \right]F\right)_{\mathbf{1}}-\frac{1}{2} M_{\Sigma} {\rm Tr} \big(( \overline{\Sigma} \tilde{\Sigma}^{c})_{\mathbf{1}} \big) - \frac{1}{2} M_F\left( \overline{F^{c}} F\right)_{\mathbf{1}} + \text{h.c.}\,,
\end{eqnarray}
where $\tilde{\eta}=i\sigma_2\eta^{*}$ and the $SU(2)_L$ triplets $\Sigma$ and $\Omega$ can be written in $2\times 2$ matrix notation as
\begin{equation}
\Omega=\left(\begin{array}{cc}
\Omega^{0}/\sqrt{2} &\Omega^{+}  \\
 \Omega^{-} &-\Omega^{0}/\sqrt{2}
\end{array}\right),~~~~
\Sigma=\left(\begin{array}{cc}
\Sigma^{0}/\sqrt{2} &\Sigma^{+}  \\
 \Sigma^{-} &-\Sigma^{0}/\sqrt{2}
\end{array}\right)
\end{equation}
and $\tilde{\Sigma}^{c}\equiv i\sigma_{2}\Sigma^{c}i\sigma_{2}$. The scalar triplet $\Omega$ is assumed to be real.

\subsection{Scalar sector}
\label{sec:scalar}

The scalar potential $\mathcal{V}$ characterizing the symmetry breaking pattern in our model can be written as
\begin{equation}
   \mathcal{V} = V(\phi) + V(\eta) + V(\Omega) + V(\phi,\eta) + V(\eta,\Omega)+ V(\phi,\Omega) \,.
\end{equation}
where
\begin{eqnarray}
V(\phi)&=& -m_{\phi}^{2}(\phi^{\dag}\phi)_{\mathbf{1}}+\frac{\lambda_{1,1}}{2}\left(\phi^{\dag}\phi\right)_{\mathbf{1}}\left(\phi^{\dag}\phi\right)_{\mathbf{1}}+\frac{\lambda_{1,2}}{2}\left(\phi^{\dag}\phi\right)_{\mathbf{1'}}\left(\phi^{\dag}\phi\right)_{\mathbf{1''}}+\frac{\lambda_{1,3}}{2}\left(\phi^{\dag}\phi\right)_{\mathbf{3}_{S}}\left(\phi^{\dag}\phi\right)_{\mathbf{3}_{S}} \nonumber\\
&~&+\frac{i\lambda_{1,4}}{2}\left(\phi^{\dag}\phi\right)_{\mathbf{3}_{S}}\left(\phi^{\dag}\phi\right)_{\mathbf{3}_{A}}+\frac{\lambda_{1,5}}{2}\left(\phi^{\dag}\phi\right)_{\mathbf{3}_{A}}\left(\phi^{\dag}\phi\right)_{\mathbf{3}_{A}}\,, \nonumber \\
V(\eta)&=&m_{\eta}^{2}(\eta^{\dag}\eta)_{\mathbf{1}}+\frac{\lambda_2}{2}\left(\eta^{\dag}\eta\right)^{2}\,,\nonumber \\
 V(\Omega)&=& -\frac{m_{\Omega}^{2}}{2}\textrm{Tr} \left(\left(\Omega^{\dag}\Omega\right)_{\mathbf{1}}\right) +\frac{\lambda_{1}^{\Omega}}{4} \textrm{Tr}\left(\left( \Omega^{\dag}\Omega \right)_{\mathbf{1}} \right) \textrm{Tr}\left(\left( \Omega^{\dag}\Omega \right)_{\mathbf{1}} \right) +\frac{\lambda_{2}^{\Omega}}{4} \textrm{Tr}\left(\left( \Omega^{\dag}\Omega \right)_{\mathbf{1'}} \right) \textrm{Tr}\left(\left( \Omega^{\dag}\Omega \right)_{\mathbf{1''}} \right) \nonumber \\
 &~&+\frac{\lambda_{3}^{\Omega}}{4} \textrm{Tr}\left(\left( \Omega^{\dag}\Omega \right)_{\mathbf{3}_{S}} \right) \textrm{Tr}\left(\left( \Omega^{\dag}\Omega \right)_{\mathbf{3}_{S}} \right) +\frac{i\lambda_{4}^{\Omega}}{4} \textrm{Tr}\left(\left( \Omega^{\dag}\Omega \right)_{\mathbf{3}_{S}} \right) \textrm{Tr}\left(\left( \Omega^{\dag}\Omega \right)_{\mathbf{3}_{A}} \right) +\frac{\lambda_{5}^{\Omega}}{4} \textrm{Tr}\left(\left( \Omega^{\dag}\Omega \right)_{\mathbf{3}_{A}} \right) \textrm{Tr}\left(\left( \Omega^{\dag}\Omega \right)_{\mathbf{3}_{A}} \right) \,,\nonumber \\
  V(\phi,\eta)&=&  \lambda_3\left(\phi^{\dag} \phi \right)_{\mathbf{1}}\left(\eta^{\dag}\eta\right)_{\mathbf{1}}  + \lambda_{4}\left( \left(\phi^{\dag}\eta\right)_{\mathbf{3}}\left(\eta^{\dag}\phi \right)_{\mathbf{3}}\right)_{\mathbf{1}} +\left[ \frac{\lambda_5}{2}\left( \left(\phi^{\dag}\eta\right)_{\mathbf{3}}\left(\phi^{\dag}\eta\right)_{\mathbf{3}}\right)_{\mathbf{1}} + h.c \right] \,,\nonumber \\
\nonumber V(\eta,\Omega)&=&\frac{\lambda_{\eta}^{\Omega}}{2}\left(\eta^{\dag}\eta \right)_{\mathbf{1}} \textrm{Tr}\left(\left(\Omega^{\dag}\Omega\right)_{\mathbf{1}}\right)\,,\\
  V(\phi,\Omega)&=& \mu_{1,1}\left( \left( \phi^{\dag}\Omega\right)_{\mathbf{3}_{S}}\phi \right)_{\mathbf{1}} + i\mu_{1,2}\left( \left( \phi^{\dag}\Omega\right)_{\mathbf{3}_{A}}\phi \right)_{\mathbf{1}} +\frac{\lambda_{\phi,1}^{\Omega}}{2}\left(\phi^{\dag}\phi \right)_{\mathbf{1}} \textrm{Tr}\left((\Omega^{\dag}\Omega)_{\mathbf{1}}\right) \nonumber\\
 &~& +\frac{\lambda_{\phi,2}^{\Omega}}{2}\left(\phi^{\dag}\phi \right)_{\mathbf{1'}} \textrm{Tr}\left((\Omega^{\dag}\Omega)_{\mathbf{1''}}\right) +\frac{\lambda_{\phi,2}^{\Omega *}}{2}\left(\phi^{\dag}\phi \right)_{\mathbf{1''}} \textrm{Tr}\left((\Omega^{\dag}\Omega)_{\mathbf{1'}}\right) +\frac{\lambda_{\phi,4}^{\Omega}}{2}\left(\phi^{\dag}\phi \right)_{\mathbf{3}_{S}} \textrm{Tr}\left((\Omega^{\dag}\Omega)_{\mathbf{3}_{S}}\right)\nonumber\\
&~&+\frac{i\lambda_{\phi,5}^{\Omega}}{2}\left(\phi^{\dag}\phi \right)_{\mathbf{3}_{S}} \textrm{Tr}\left((\Omega^{\dag}\Omega)_{\mathbf{3}_{A}}\right) +\frac{i\lambda_{\phi,6}^{\Omega}}{2}\left(\phi^{\dag}\phi \right)_{\mathbf{3}_{A}} \textrm{Tr}\left((\Omega^{\dag}\Omega)_{\mathbf{3}_{S}}\right) +\frac{\lambda_{\phi,7}^{\Omega}}{2}\left(\phi^{\dag}\phi \right)_{\mathbf{3}_{A}} \textrm{Tr}\left((\Omega^{\dag}\Omega)_{\mathbf{3}_{A}}\right)  \,.
\end{eqnarray}
Here only $\lambda_{\phi,2}^{\Omega}$ is complex and all of other couplings are real, including $\lambda_5$, taken to be real without any loss of generality~\cite{Ma:2006km,Hirsch:2013ola}.
We assume the scalar fields have the simple vacuum alignment
\begin{equation}
\label{eq:vac_alignment}\langle\phi \rangle=\begin{pmatrix}
1 \\
1 \\
1
\end{pmatrix} v_{\phi},~~~~
\langle\Omega\rangle=\begin{pmatrix}
1 \\
0 \\
0
\end{pmatrix} v_{\Omega},~~~~
\langle\eta\rangle=0\,.
\end{equation}
The vacuum expectation values of $\phi$ and $\Omega$ break the $A_4$ flavor symmetry down to the subgroups $Z^T_3$ and $Z^S_2$ respectively, where the superscripts denote the generator of the subgroups.
This commonly used alignment can be achieved by forbidding the cross terms between $\phi$ and $\Omega$.
This may be done by appealing to supersymmetry~\cite{Babu:2002dz,Altarelli:2005yx} or extra dimensions~\cite{Altarelli:2005yp}.
In fact our desired alignment is exactly realized in~\cite{Altarelli:2005yp}.
With the vacuum alignment given in Eq.~\eqref{eq:vac_alignment}, we find that the minimization conditions are
\begin{eqnarray}\nonumber
\frac{\partial \mathcal{V}}{\partial \phi_{1}^{*}}&=&\frac{v_{\phi}}{2\sqrt{2}}\left[-2m_{\phi}^{2}+(3\lambda_{1,1}+4\lambda_{1,3})|v_{\phi}|^{2}+(\lambda_{\phi,1}^{\Omega}+2\text{Re}(\lambda_{\phi,2}^{\Omega}))|v_{\Omega}|^{2}\right]=0\,,\\
\nonumber \frac{\partial \mathcal{V}}{\partial \phi_{2}^{*}}&=&\frac{v_{\phi}}{2\sqrt{2}}\left[-2m_{\phi}^{2}-\sqrt{2}(\mu_{1,1}-i\mu_{1,2})v_{\Omega}+(3\lambda_{1,1}+4\lambda_{1,3})|v_{\phi}|^{2}+(\lambda_{\phi,1}^{\Omega}+\sqrt{3}\text{Im}(\lambda_{\phi,2}^{\Omega})-\text{Re}(\lambda_{\phi,2}^{\Omega}))|v_{\Omega}|^{2}\right]=0\,,\\
\nonumber\frac{\partial \mathcal{V}}{\partial \phi_{3}^{*}}&=&\frac{v_{\phi}}{2\sqrt{2}}\left[ -2m_{\phi}^{2}-\sqrt{2}(\mu_{1,1}+i\mu_{1,2})v_{\Omega}+(3\lambda_{1,1}+4\lambda_{1,3})|v_{\phi}|^{2}+(\lambda_{\phi,1}^{\Omega}-\sqrt{3}\text{Im}(\lambda_{\phi,2}^{\Omega})-\text{Re}(\lambda_{\phi,2}^{\Omega}))|v_{\Omega}|^{2}\right]=0\,,\\
\nonumber \frac{\partial \mathcal{V}}{\partial \Omega^{0}_{1}}&=&-m_{\Omega}^{2}v_{\Omega}+(\lambda_{1}^{\Omega}+\lambda_{2}^{\Omega})v_{\Omega}^{3}+\frac{1}{2}(3\lambda_{\phi ,1}^{\Omega}v_{\Omega}-\sqrt{2}\mu_{1,1})|v_{\phi}|^{2}=0\,,\\
 \nonumber  \frac{\partial \mathcal{V}}{\partial \Omega^{0}_{2}}&=&(\lambda_{\phi ,4}^{\Omega}v_{\Omega}-\frac{\mu_{1,1}}{\sqrt{2}})|v_{\phi}|^{2}=0\,,\\
\frac{\partial \mathcal{V}}{\partial \Omega^{0}_{3}}&=&(\lambda_{\phi ,4}^{\Omega}v_{\Omega}-\frac{\mu_{1,1}}{\sqrt{2}})|v_{\phi}|^{2}=0\,.
\end{eqnarray}
From above equations, we find that the non-trivial solutions can be achieved if the following relations among the parameters are satisfied
\begin{equation}
 \mu_{1,1}=-\frac{3\sqrt{2}}{2}\text{Re}(\lambda_{\phi,2}^{\Omega})v_{\Omega}\,,\quad
 \mu_{1,2}=\text{Im}(\lambda_{\phi,2}^{\Omega})=0\,,\quad
\lambda_{\phi ,4}^{\Omega}=-\frac{3}{2}\text{Re}(\lambda_{\phi,2}^{\Omega})\,.
\end{equation}
Under such assumptions, the minimization conditions are simplified into
\begin{eqnarray}\nonumber
  &&\frac{\partial \mathcal{V}}{\partial \phi_{1}^{*}}=\frac{\partial \mathcal{V}}{\partial \phi_{2}^{*}}=\frac{\partial \mathcal{V}}{\partial \phi_{3}^{*}}=\frac{v_{\phi}}{2\sqrt{2}}\left( -2m_{\phi}^{2}+(3\lambda_{1,1}+4\lambda_{1,3})|v_{\phi}|^{2}+(\lambda_{\phi,1}^{\Omega}+2\text{Re}(\lambda_{\phi,2}^{\Omega}))v_{\Omega}^{2}\right)=0\,,\\
  &&\frac{\partial \mathcal{V}}{\partial \Omega^{0}_{1}}=v_{\Omega}\left(-m_{\Omega}^{2}+(\lambda_{1}^{\Omega}+\lambda_{2}^{\Omega})v_{\Omega}^{2}+\frac{3}{2}(\lambda_{\phi ,1}^{\Omega}+\text{Re}(\lambda_{\phi,2}^{\Omega}))|v_{\phi}|^{2}\right)=0\,,~~\frac{\partial \mathcal{V}}{\partial \Omega^{0}_{2}}=\frac{\partial \mathcal{V}}{\partial \Omega^{0}_{3}}=0\,.
\end{eqnarray}
We assume that the cross terms $V(\phi, \Omega)$ between $\phi$ and $\Omega$ can be forbidden.
In this case, the minimization conditions can be written as
\begin{eqnarray}\nonumber
  &&\frac{\partial \mathcal{V}}{\partial \phi_{1}^{*}}=\frac{\partial \mathcal{V}}{\partial \phi_{2}^{*}}=\frac{\partial \mathcal{V}}{\partial \phi_{3}^{*}}=\frac{v_{\phi}}{2\sqrt{2}}\left( -2m_{\phi}^{2}+(3\lambda_{1,1}+4\lambda_{1,3})|v_{\phi}|^{2}\right)=0\,,\\
  &&\frac{\partial \mathcal{V}}{\partial \Omega^{0}_{1}}=-m_{\Omega}^{2}v_{\Omega}+(\lambda_{1}^{\Omega}+\lambda_{2}^{\Omega})v_{\Omega}^{3}=0\,,~~\frac{\partial \mathcal{V}}{\partial \Omega^{0}_{2}}=\frac{\partial \mathcal{V}}{\partial \Omega^{0}_{3}}=0\,,
\end{eqnarray}
then we can obtain
\begin{eqnarray}
  |v_{\phi}|^{2}=\frac{2m_{\phi}^{2}}{3\lambda_{1,1}+4\lambda_{1,3}}\,,\quad
  v_{\Omega}^{2}=\frac{m_{\Omega}^{2}}{\lambda_{1}^{\Omega}+\lambda_{2}^{\Omega}}\,.
\end{eqnarray}
The mass matrix of the neutral scalars in the basis $(h^{0}_{1}, h^{0}_{2}, h^{0}_{3},\Omega_{1}^{0},\Omega_{2}^{0},\Omega_{3}^{0})$ are
\begin{equation}
M_{h}^{2}=\left( \begin{array}{cc} M_{h^{0}}^{2} ~&~ 0_{3\times 3} \\
0_{3\times 3} ~&~ M_{\Omega^{0}}^{2} \\ \end{array} \right)\,,
\end{equation}
where
\begin{eqnarray}\nonumber
M_{h^{0}}^{2}&=&\frac{v_{\phi}^{2}}{2}\left( \begin{array}{ccc} 2(\lambda_{1,2}+\lambda_{1,2}) ~& 2\lambda_{1,1}-\lambda_{1,2}+4\lambda_{1,3} ~& 2\lambda_{1,1}-\lambda_{1,2}+4\lambda_{1,3} \\
2\lambda_{1,1}-\lambda_{1,2}+4\lambda_{1,3} ~& 2(\lambda_{1,2}+\lambda_{1,2}) ~& 2\lambda_{1,1}-\lambda_{1,2}+4\lambda_{1,3} \\
2\lambda_{1,1}-\lambda_{1,2}+4\lambda_{1,3} ~& 2\lambda_{1,1}-\lambda_{1,2}+4\lambda_{1,3} ~& 2(\lambda_{1,2}+\lambda_{1,2}) \\ \end{array} \right) \,,\\
M_{\Omega^{0}}^{2}&=&\text{diag}\left(2(\lambda_{1}^{\Omega}+\lambda_{2}^{\Omega})v_{\Omega}^{2}, ~(-\frac{3}{2}\lambda_{2}^{\Omega}+2\lambda_{3}^{\Omega})v_{\Omega}^{2},~ (-\frac{3}{2}\lambda_{2}^{\Omega}+2\lambda_{3}^{\Omega})v_{\Omega}^{2}\right)\,.
\end{eqnarray}
The mass matrix for the charged scalars in the basis $(\phi^{\pm}_{1,2,3},\Omega^{\pm}_{1,2,3})$ is
\begin{equation}
M_{H^{\pm}}^{2}=\left( \begin{array}{cc} M_{\phi^{\pm}}^{2} ~&~ 0_{3\times 3} \\
0_{3\times 3} ~&~ M_{\Omega^{\pm}}^{2} \\ \end{array} \right)\,,
\end{equation}
with
\begin{eqnarray}\nonumber
M_{\phi^{\pm}}^{2}&=&\frac{v_{\phi}^{2}}{2}\left( \begin{array}{ccc} -4\lambda_{1,3} & 2\lambda_{1,3}+i\lambda_{1,4} & 2\lambda_{1,3}-i\lambda_{1,4}\\
2\lambda_{1,3}-i\lambda_{1,4} & -4\lambda_{1,3} & 2\lambda_{1,3}+i\lambda_{1,4} \\
2\lambda_{1,3}+i\lambda_{1,4} & 2\lambda_{1,3}-i\lambda_{1,4} & -4\lambda_{1,3} \\ \end{array} \right)\,,\\
M_{\Omega^{\pm}}^{2}&=&\text{diag}\left( 0,~ -\frac{3}{2}\lambda^{\Omega}_{2}v_{\Omega}^{2}, ~ -\frac{3}{2}\lambda^{\Omega}_{2}v_{\Omega}^{2}\right)\,.
\end{eqnarray}
Notice that the massless $\Omega^{\pm}_{1}$ is absorbed by the $W$ boson. The $W$ and $Z$ boson masses are
\begin{equation}
   m_{W}^{2}=\frac{1}{4}g^{2}(3v_{\phi}^{2}+4 v_{\Omega}^{2})\,,\quad m_{Z}^{2}=\frac{3}{4}(g^{2}+g'^{2})v_{\phi}^{2}\,.
 \end{equation}
Concerning the dark scalar $\eta$, taking the pattern of scalar vacuum expectation values in Eq.~\eqref{eq:vac_alignment}, we can read out the mass eigenvalues of the $\eta$ scalars.
The masses of the real and imaginary parts of the neutral $\eta^0=(\eta_R+i\eta_I)/\sqrt{2}$ are
\begin{eqnarray}\nonumber
m_{\eta_R}^2 &=& m_{\eta}^2 + \frac{3}{2}\left(\lambda_3 + \lambda_4 + \lambda_5 \right) v_{\phi}^2 + \frac{1}{2}\lambda^\eta_{\Omega} v_\Omega^2  \, ,\label{eq:etaRmass} \\
m_{\eta_I}^2 &=& m_{\eta}^2  + \frac{3}{2}\left(\lambda_3 + \lambda_4 - \lambda_5 \right) v_{\phi}^2 + \frac{1}{2}\lambda^\eta_{\Omega} v_\Omega^2 \, ,\label{eq:etaImass}
\end{eqnarray}
One sees that, in the limit $\lambda_5 \to 0$ the masses of the scalars $(\eta_{R},\eta_{I})$ are degenerate, a characteristic feature of scotogenic models. The mass of charged $\eta$ bosons is
\begin{equation}
  m^{2}_{\eta^{\pm}}=m_{\eta}^{2}+\frac{3}{2}\lambda_{3}v_{\phi}^{2}+\frac{1}{2}\lambda_{\eta}^{\Omega}v_{\Omega}^{2}\,.
\end{equation}


\subsection{Charged lepton sector}
\label{sec:Charged_lepton}


The Yukawa terms responsible for generating the charged lepton masses are
\begin{eqnarray}
\nonumber\mathcal{L}_{\ell}&=&-y_e (\overline{L} \phi)_{\mathbf{1}} e_R-y_{\mu} (\overline{L} \phi)_{\mathbf{1''}} \mu_R-y_{\tau} (\overline{L} \phi)_{\mathbf{1'}} \tau_R\,,\\
\label{eq:charged-lepton-Yukawa}&=&-y_e(\overline{e_L}\phi_{1}+\overline{\mu_L}\phi_{2}+\overline{\tau_L}\phi_{3})e_R-y_{\mu}(\overline{e_L}\phi_{1}+\omega\overline{\mu_L}\phi_{2}+\omega^2\overline{\tau_L}\phi_{3})\mu_R-y_{\tau}(\overline{e_L}\phi_{1}+\omega^2\overline{\mu_L}\phi_{2}+\omega\overline{\tau_L}\phi_{3})\tau_R
\end{eqnarray}
For the alignment of $\phi$ in Eq.~\eqref{eq:vac_alignment}, one can read out the charged lepton mass matrix as follows
\begin{equation}
M_{\ell}=\begin{pmatrix}
y_e   &   y_{\mu}   & y_{\tau}  \\
y_{e} & \omega y_{\mu}  & \omega^2y_{\tau} \\
y_e  &  \omega^2 y_{\mu} & \omega y_{\tau} \\
\end{pmatrix}v_{\phi}=\frac{1}{\sqrt{3}} \begin{pmatrix}
 1 &   1   & 1  \\
1 & \omega   & \omega^2 \\
1  &  \omega^2 & \omega  \\
\end{pmatrix}\text{diag}(\sqrt{3}y_e v_{\phi}, \sqrt{3}y_{\mu} v_{\phi}, \sqrt{3}y_{\tau} v_{\phi})
\end{equation}
Therefore the charged lepton mass matrix $M_{\ell}$ is diagonalized by the unitary transformation
\begin{equation}
U_{\ell}=\frac{1}{\sqrt{3}} \begin{pmatrix}
 1 &   1   & 1  \\
1 & \omega   & \omega^2 \\
1  &  \omega^2 & \omega  \\
\end{pmatrix}\,,
\label{eq:Ul}
\end{equation}
so that $U^{\dagger}_{\ell}M_{\ell}M^{\dagger}_{\ell}U_{\ell}$ is diagonal and positive.

\subsection{Dark fermionic sector}
\label{sec:dark-ferm-sect}

In addition to their bare mass terms, the dark Majorana fermions $F$ and $\Sigma$ have Yukawa couplings to $\Omega$, which also contribute to their masses,
\begin{equation}
\mathcal{L}_{F\Sigma} = -Y_{\Omega,1}\left( {\textrm{Tr}} \left[\left(\overline{\Sigma}\Omega\right)_{\mathbf{3}_{S}} \right]F\right)_{\mathbf{1}}-Y_{\Omega,2}\left( {\textrm{Tr}} \left[\left(\overline{\Sigma}\Omega\right)_{\mathbf{3}_{A}} \right]F\right)_{\mathbf{1}}-\frac{1}{2} M_{\Sigma} {\rm Tr} \big(( \overline{\Sigma} \tilde{\Sigma}^{c})_{\mathbf{1}} \big) - \frac{1}{2} M_F\left( \overline{F^{c}} F\right)_{\mathbf{1}} +\text{h.c.}\,.
\end{equation}
Given the vacuum expectation values of scalars in Eq.~\eqref{eq:vac_alignment} one can read out the mass matrix of the dark fermions $F$ and $\Sigma^0$.
In the basis of $\left( \Sigma_{0}^{c}, F \right)$, their Majorana mass terms are defined as
\begin{equation}
-\frac{1}{2}(\overline{\Sigma^0_{1}},\overline{F^{c}_{1}},\overline{\Sigma^0_{2}},\overline{F^{c}_{3}},\overline{\Sigma^0_{3}},\overline{F^{c}_{2}})M_{\chi}\begin{pmatrix} \Sigma_{1}^{0c} \\ F_{1} \\ \Sigma_{2}^{0c} \\ F_{3} \\ \Sigma_{3}^{0c} \\ F_{2} \end{pmatrix} \,,
\end{equation}
where the $6\times 6$ Majorana mass matrix takes the form
\begin{equation}
  \label{eq:M_chi_1}
M_\chi = \left(\begin{array}{cccccc} M_\Sigma & 0 & 0 & 0 & 0 & 0 \\
0 & M_{F} & 0 & 0 & 0 & 0 \\
0 &  0 & M_\Sigma  & (Y_{\Omega,1}-Y_{\Omega,2})v_{\Omega} & 0 & 0 \\
0 &  0 & (Y_{\Omega,1}-Y_{\Omega,2})v_{\Omega} & M_{F} & 0 & 0 \\
0 &  0 & 0 & 0 & M_\Sigma  & (Y_{\Omega,1}+Y_{\Omega,2})v_{\Omega} \\
0 &  0 & 0 & 0 & (Y_{\Omega,1}+Y_{\Omega,2})v_{\Omega} & M_{F} \\
\end{array}\right)  \, .
\end{equation}
Notice that, from the electroweak precision tests, i.e. restrictions due to the $\rho$ parameter~\cite{Tanabashi:2018oca} one expects a small mixing between the two dark sub-sectors.
The symmetric complex $6\times 6$ matrix $M_{\chi}$ in Eq.\eqref{eq:M_chi_1} can be diagonalized~\cite{Schechter:1980gr} by a $6\times 6$ block-diagonal matrix $V$~\cite{Schechter:1981cv}
\begin{equation}
V^{T}M_{\chi}V=\text{diag}(m_{\chi^{0}_{1}},m_{\chi^{0}_{2}},m_{\chi^{0}_{3}},m_{\chi^{0}_{4}},m_{\chi^{0}_{5}},m_{\chi^{0}_{6}})\,,
\end{equation}
with
\begin{equation}
V = \left(\begin{array}{cccccc}1 & 0 & 0 & 0 & 0 & 0 \\
0 & 1 & 0 & 0 & 0 & 0 \\
0 & 0 &\cos\theta e^{i(\phi_{1}+\varrho_{1})/2} & \sin\theta e^{i(\phi_{1}+\sigma_{1})/2} ~& 0~ & ~0 \\
0 & 0 &  -\sin\theta e^{i(-\phi_{1}+\varrho_{1})/2} & \cos\theta e^{i(-\phi_{1}+\sigma_{1})/2} ~& 0~ & ~0 \\
0 & 0 & 0 & 0 ~& ~\cos\alpha e^{i(\phi_{2}+\varrho_{2})/2} & \sin\alpha e^{i(\phi_{2}+\sigma_{2})/2}\\
0 & 0 & 0 & 0 ~& ~-\sin\alpha e^{i(-\phi_{2}+\varrho_{2})/2}& \cos\alpha e^{i(-\phi_{2}+\sigma_{2})/2} \\ \end{array}\right)\,,
\end{equation}
with the rotation angles $\theta$ and $\alpha$ satisfying
\begin{equation}
\tan2\theta=\frac{\Delta_{34}}{M_{F}^{2}-M_{\Sigma}^{2}}\,,\qquad
\tan2\alpha=\frac{\Delta_{56}}{M_{F}^{2}-M_{\Sigma}^{2}}\,,
\end{equation}
with
\begin{eqnarray}
  \Delta_{34} = 2Y_{-}\sqrt{M_{\Sigma}^{2}+M_{F}^{2}+2M_{\Sigma}M_{F}\cos2\phi_{34}}\,,\quad
  \Delta_{56} = 2Y_{+}\sqrt{M_{\Sigma}^{2}+M_{F}^{2}+2M_{\Sigma}M_{F}\cos2\phi_{56}}\,.
\end{eqnarray}  
Here we have defined
\begin{eqnarray}\nonumber
&&Y_{-}\equiv |(Y_{\Omega,1}-Y_{\Omega,2})v_{\Omega}|\,,\quad \phi_{34}\equiv \text{arg}((Y_{\Omega,1}-Y_{\Omega,2})v_{\Omega})\,,\\
&&Y_{+}\equiv |(Y_{\Omega,1}+Y_{\Omega,2})v_{\Omega}|\,,\quad\phi_{56}\equiv \text{arg}((Y_{\Omega,1}+Y_{\Omega,2})v_{\Omega})\,.
\end{eqnarray}
The eigenvalues $M_{\chi_{1,2,3,4,5,6}^{0}}$ are given as
\begin{eqnarray}\nonumber
&&m_{\chi_{1}^{0}}=M_{\Sigma}\,,~~~ m_{\chi_{2}^{0}}=M_{F}\,,\\ \nonumber
&&m_{\chi_{3}^{0}}^{2}=\frac{1}{2}\left( M_{\Sigma}^{2}+M_{F}^{2}+2Y_{-}^{2}-\sqrt{(M_{F}^{2}-M_{\Sigma}^{2})^{2}+\Delta_{34}^{2}}\right)\,,\\ \nonumber
&&m_{\chi_{4}^{0}}^{2}=\frac{1}{2}\left( M_{\Sigma}^{2}+M_{F}^{2}+2Y_{-}^{2}+\sqrt{(M_{F}^{2}-M_{\Sigma}^{2})^{2}+\Delta_{34}^{2}}\right)\,,\\ \nonumber
&&m_{\chi_{5}^{0}}^{2}=\frac{1}{2}\left( M_{\Sigma}^{2}+M_{F}^{2}+2Y_{+}^{2}-\sqrt{(M_{F}^{2}-M_{\Sigma}^{2})^{2}+\Delta_{56}^{2}}\right)\,,\\ 
&&m_{\chi_{6}^{0}}^{2}=\frac{1}{2}\left( M_{\Sigma}^{2}+M_{F}^{2}+2Y_{+}^{2}+\sqrt{(M_{F}^{2}-M_{\Sigma}^{2})^{2}+\Delta_{56}^{2}}\right)\label{mchi2}\,,
\end{eqnarray}
In the limit $v_{\Omega}\ll M_{\Sigma}, M_{F}$, which means $Y_{\pm}^{2}\ll M_{\Sigma}^{2}, M_{F}^{2}$, then the masses of fermions $\chi^{0}_{1,2,3,4,5,6}$ are approximately degenerate as
\begin{eqnarray}
\nonumber&&m_{\chi_{1}^{0}}=M_{\Sigma}\,,~~~m_{\chi_{2}^{0}}=M_{F}\,,\\
\nonumber&&m_{\chi_{3}^{0}}^{2}=M^2_{\Sigma}-2Y^2_{-}\frac{M_{\Sigma}(M_{\Sigma}+M_F\cos2\phi_{34})}{M^2_F-M^2_{\Sigma}},~~m_{\chi_{4}^{0}}^{2}=M^2_{F}+2Y^2_{-}\frac{M_{F}(M_{F}+M_{\Sigma}\cos2\phi_{34})}{M^2_F-M^2_{\Sigma}}\,,\\ &&m_{\chi_{5}^{0}}^{2}=M^2_{\Sigma}-2Y^2_{+}\frac{M_{\Sigma}(M_{\Sigma}+M_F\cos2\phi_{56})}{M^2_F-M^2_{\Sigma}},~~m_{\chi_{6}^{0}}^{2}=M^2_{F}+2Y^2_{+}\frac{M_{F}(M_{F}+M_{\Sigma}\cos2\phi_{56})}{M^2_F-M^2_{\Sigma}}\,,
\end{eqnarray}
for $M_F>M_{\Sigma}$. The expressions of $m_{\chi_{3}^{0}}$ and $m_{\chi_{5}^{0}}$ are exchanged with those of $m_{\chi_{4}^{0}}$ and $m_{\chi_{6}^{0}}$ respectively for $M_F<M_{\Sigma}$.
The Majorana mass eigenstates $\chi^{0}_{1,2,3,4,5,6}$ are related with $\Sigma^{c}_{0}$ and $F$ by the unitary transformation $V$ as follow
\begin{equation}
\label{eq:relation_eigenstates_A4_1}
\begin{pmatrix} \chi^{0}_{1} \\ \chi^{0}_{2} \\ \chi^{0}_{3} \\ \chi^{0}_{4}\\ \chi^{0}_{5} \\ \chi^{0}_{6}\end{pmatrix}=V^{\dag} \begin{pmatrix} \Sigma_{1}^{0c} \\ F_{1} \\ \Sigma_{2}^{0c} \\ F_{3} \\ \Sigma_{3}^{0c} \\ F_{2} \end{pmatrix} \,,~~\quad~~  \begin{pmatrix} \Sigma_{1}^{0c} \\ F_{1} \\ \Sigma_{2}^{0c} \\ F_{3} \\ \Sigma_{3}^{0c} \\ F_{2} \end{pmatrix} =V \begin{pmatrix} \chi^{0}_{1} \\ \chi^{0}_{2} \\ \chi^{0}_{3} \\ \chi^{0}_{4}\\ \chi^{0}_{5} \\ \chi^{0}_{6}\end{pmatrix}\,.
\end{equation}

\section{Neutrino mass}
\label{sec:neutrino_mass}

\begin{figure}[t!]
\centering
\includegraphics[scale=0.45]{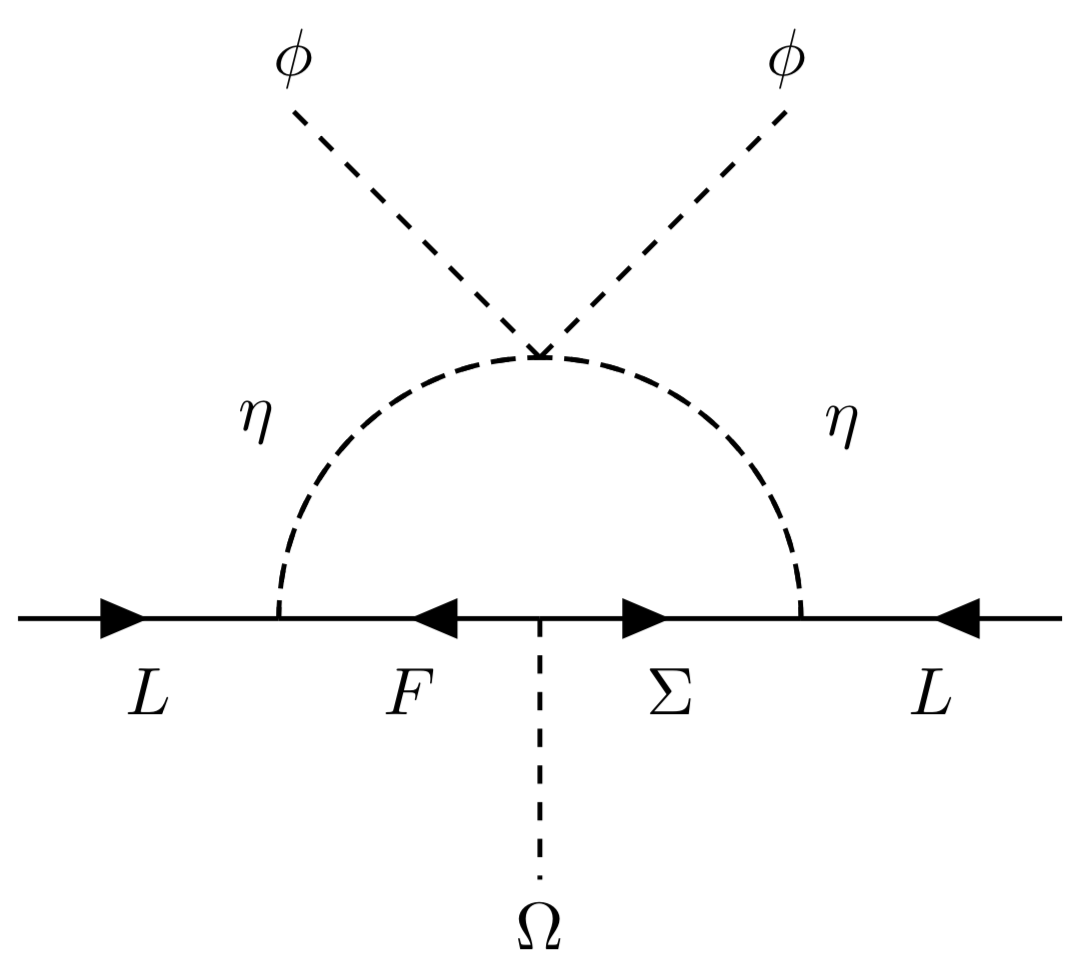}
\caption{\label{fig:loop} Feynman diagram for scotogenic neutrino mass generation. }
\end{figure}

Neutrino masses are generated at the one-loop level, as shown in Fig.~\ref{fig:loop}.
This mechanism is simply a flavour extension of the original singlet-triplet scotogenic model~\cite{Hirsch:2013ola,Merle:2016scw,Diaz:2016udz,Choubey:2017yyn,Restrepo:2019ilz,Avila:2019hhv}.
Notice that, in contrast to the dark scalar $\eta$, the dark fermions transform as $A_4$ triplets.
We stress that this figure actually includes all possible 1-loop diagrams, since $\eta^0 = (\eta_{R} + i\eta_{I})/\sqrt{2}$ and all six intermediate dark fermions are included.
The interactions contributing to neutrino mass generation arise from the terms $-Y_{F}\left(\overline{L}F\right)_{\mathbf{1}}\tilde{\eta} - Y_{\Sigma}(\overline{L}\tilde{\Sigma}^{c})_{\mathbf{1}}\tilde{\eta}$ in Eq.~\eqref{eq:Lagrangian-fermion}, we have 
\begin{equation}
\mathcal{L}_{\nu}=-\overline{\nu_{e}}(Y_{F}F_{1}+\frac{Y_{\Sigma}}{\sqrt{2}}\Sigma_{1}^{0c})\eta^{*}_{0} -\overline{\nu_{\mu}}(Y_{F}F_{2}+\frac{Y_{\Sigma}}{\sqrt{2}}\Sigma_{2}^{0c})\eta^{*}_{0} -\overline{\nu_{\tau}}(Y_{F}F_{3}+\frac{Y_{\Sigma}}{\sqrt{2}}\Sigma_{3}^{0c})\eta^{*}_{0}+\text{h.c.}\,.
\end{equation}
In the mass eigenstate of dark Majorana fermions, $\mathcal{L}_{\nu}$ is of the following form
\begin{equation}
\label{eq:vertex}\mathcal{L}_{\nu}= -\frac{1}{\sqrt{2}}h_{\alpha i}\overline{\nu_{\alpha}}\eta_{R}\chi_{i}^{0}  + \frac{i}{\sqrt{2}}h_{\alpha i}\overline{\nu_{\alpha}}\eta_{I}\chi_{i}^{0}
-\frac{1}{\sqrt{2}}h_{\alpha i}^{*}\overline{\chi_{i}^{0}}\eta_{R}\nu_{\alpha} -\frac{i}{\sqrt{2}}h_{\alpha i}^{*} \overline{\chi_{i}^{0}}\eta_{I}\nu_{\alpha} \,,
\end{equation}
where we have defined
\begin{equation}
h=\left( \begin{array}{cccccc} \frac{Y_{\Sigma}}{\sqrt{2}} & Y_{F} & 0 & 0 & 0 & 0 \\
0 & 0 & \frac{Y_{\Sigma}}{\sqrt{2}}  & 0 & 0 & Y_{F} \\
0 & 0 & 0 & Y_{F} & \frac{Y_{\Sigma}}{\sqrt{2}} & 0 \\ \end{array} \right)V\,.
\end{equation}
We find the neutrino masses radiatively generated at one-loop level are given by  %
\begin{eqnarray}
  \nonumber
  i(\mathcal{M}_{\nu})_{\alpha\beta}
  &=&\sum_{i}\frac{1}{2} \int \frac{d^{4} k }{2\pi ^{4}}\left( \frac{-i h_{\alpha i} h_{\beta i}}{k^{2}-m_{\eta_{R}}^{2}} + \frac{i h_{\alpha i} h_{\beta i}}{k^{2}-m_{\eta_{I}}^{2}}\right) \frac{i (\cancel{k}+m_{\chi_{i}^{0}})}{k^{2}-m_{\chi_{i}^{0}}^{2}}\\
  \nonumber
&=&i\sum_{i}\frac{h_{\alpha i} h_{\beta i}}{32\pi^{2}}m_{\chi^{0}_{i}}\left( -\frac{m^{2}_{\eta_{R}}\text{ln}(\frac{m^{2}_{\eta_{R}}}{m^{2}_{\chi^{0}_{i}}})}{m^{2}_{\eta_{R}}-m^{2}_{\chi^{0}_{i}}} + \frac{m^{2}_{\eta_{I}}\text{ln}(\frac{m^{2}_{\eta_{I}}}{m^{2}_{\chi^{0}_{i}}})}{m^{2}_{\eta_{I}}-m^{2}_{\chi^{0}_{i}}}\right)\\
&=&i\sum_{i}\frac{h_{\alpha i} h_{\beta i}}{32\pi^{2}}m^{\nu}_i\,,
\end{eqnarray}
where
\begin{equation}
m^{\nu}_i\equiv m_{\chi^{0}_{i}}\left[ -\frac{m^{2}_{\eta_{R}}\text{ln}(\frac{m^{2}_{\eta_{R}}}{m^{2}_{\chi^{0}_{i}}})}{m^{2}_{\eta_{R}}-m^{2}_{\chi^{0}_{i}}} + \frac{m^{2}_{\eta_{I}}\text{ln}(\frac{m^{2}_{\eta_{I}}}{m^{2}_{\chi^{0}_{i}}})}{m^{2}_{\eta_{I}}-m^{2}_{\chi^{0}_{i}}}\right]\,.
\end{equation}
Thus the neutrino mass matrix can be written in matrix form as
\begin{eqnarray}
  \mathcal{M_{\nu}}=\frac{1}{32\pi^{2}}h.\text{diag}(m^{\nu}_{1},m^{\nu}_{2},m^{\nu}_{3},m^{\nu}_{4},m^{\nu}_{5},m^{\nu}_{6}).h^{T}\,.
\end{eqnarray}
Hence one finds a very simple pattern, namely,
\begin{equation}
  \mathcal{M_{\nu}}=\left( \begin{array}{ccc}
                                                                                      a & 0 & 0 \\
                                                                                      0 & b & c \\
                                                                                      0 & c & d \\ \end{array} \right)\,.
\end{equation}
where
\begin{eqnarray}\nonumber
&&a= \frac{1}{32\pi^{2}}\left[ \frac{Y_{\Sigma}^{2}}{2}m^{\nu}_{1} + Y_{F}^{2}m^{\nu}_{2} \right]\,,\\
\nonumber&&b=\frac{1}{32\pi^{2}}\left[ \frac{Y_{\Sigma}^{2}}{2}\left( e^{i(\varrho_{1}+\phi_{1})}m^{\nu}_{3}\cos^{2}\theta +e^{i(\sigma_{1}+\phi_{1})} m^{\nu}_{4}\sin^{2}\theta\right) +Y_{F}^{2}\left( e^{i(\varrho_{2}-\phi_{2})} m^{\nu}_{5}\sin^{2}\alpha +e^{i(\sigma_{2}-\phi_{2})}m^{\nu}_{6}\cos^{2}\alpha\right) \right] \,,\\
\nonumber&&c=\frac{Y_{F}Y_{\Sigma}}{64\sqrt{2}\pi^{2}}\left[ (e^{i\sigma_{1}}m^{\nu}_{4}-e^{i\varrho_{1}}m^{\nu}_{3})\sin2\theta + (e^{i\sigma_{2}}m^{\nu}_{6}-e^{i\varrho_{2}}m^{\nu}_{5})\sin2\alpha\right]\,,\\
&&d=\frac{1}{32\pi^{2}}\left[ Y_{F}^{2}\left( e^{i(\varrho_{1}-\phi_{1})}m^{\nu}_{3}\sin^{2}\theta+e^{i(\sigma_{1}-\phi_{1})}m^{\nu}_{4}\cos^{2}\theta\right) + \frac{Y_{\Sigma}^{2}}{2}\left(e^{i(\varrho_{2}+\phi_{2})}m^{\nu}_{5}\cos^{2}\alpha+e^{i(\sigma_{2}+\phi_{2})}m^{\nu}_{6}\sin^{2}\alpha \right) \right]\,.
\end{eqnarray}
Notice the simple structure of the neutrino mass matrix, with two vanishing entries. Since $\mathcal{M}_{\nu}$ is a symmetric matrix, it can be diagonalized as
\begin{equation}
U_{\nu}^{\dagger}\mathcal{M}_{\nu}U_{\nu}^{*}=\text{diag}(m_{1},m_{2},m_{3})\,,
\end{equation}
where $U_{\nu}$ can be generally denoted as
\begin{equation}
  U_{\nu}=\left(\begin{array}{ccc} 1 & 0 & 0 \\ 0 & \cos\theta_{\nu} & \sin\theta_{\nu} e^{i\delta_{\nu}} \\ 0 & -\sin\theta_{\nu} e^{-i\delta_{\nu}} & \cos\theta_{\nu} \\ \end{array} \right)\,.
  \label{eq:U_nu}
\end{equation}
The parameters $\theta_{\nu}$ and $\delta_{\nu}$ satisfy
\begin{equation}
  \tan 2\theta_{\nu}=\frac{\Delta}{|d|^{2}-|b|^{2}}\,,
  \quad  \delta_{\nu}=\text{arg}(bc^{*}+cd^{*})\,.
\end{equation}
with
\begin{equation}
  \Delta = 2|c|\sqrt{|b|^{2}+|d|^{2}+2|b||d|\cos\left( \text{arg}(b)+\text{arg}(d)-2\text{arg}(c)\right)}\,.
\end{equation}
The light neutrino masses $m_{i}$ are given as
\begin{eqnarray}\nonumber
  &&m_{1}=a\,,\\ \nonumber
  &&m_{2}^{2}=\frac{1}{2}\left( |b|^{2}+|d|^{2}+2|c|^{2}-\mathcal{S}\sqrt{(|d|^{2}-|b|^{2})^{2}+\Delta^{2}} \right)\,,\\
  &&m_{3}^{2}=\frac{1}{2}\left( |b|^{2}+|d|^{2}+2|c|^{2}+\mathcal{S}\sqrt{(|d|^{2}-|b|^{2})^{2}+\Delta^{2}} \right)\,.
\end{eqnarray}
where $\mathcal{S}=\text{sign}\left((|d|^{2}-|b|^{2})\cos2\theta_{\nu}\right)$. And $\mathcal{S}$ equals to $1$ $(-1)$ corresponds to the ordering of neutrino masses is normal (inverted).
At this point we stress that the special structure of the light neutrino mass matrix.
The parameter $a$ is real, and the other three parameters $b$, $c$, $d$ are complex.
Altogether, we are left with 7 real input parameters: $a$, $\text{Re}(b)$, $\text{Im}(b)$, $\text{Re}(c)$, $\text{Im}(c)$, $\text{Re}(d)$ and $\text{Im}(d)$ to explain 9 observables including 3 neutrino masses, 3 mixing angles and three CP violation phases.
Therefore we expect to have $9-7=2$ predictions. This will be made explicit in Sec.~\ref{sec:lepton_mixing}.

\section{The lepton mixing matrix}
\label{sec:lepton_mixing}

The lepton mixing matrix characterizing the charged current weak interaction arises from the mismatch in the diagonalizations of the charged and neutral lepton mass matrices~\cite{Schechter:1980gr}.
In our model the corresponding diagonalizing matrices are given in Eqs.~(\ref{eq:Ul}) and (\ref{eq:U_nu}), respectively.
Combining the results of $U_{l}$ and $U_{\nu}$ we obtain the general form of lepton mixing matrix as
\begin{eqnarray}
  \label{eq:Umix}
  U_{}=U_{l}^{\dag}U_{\nu}=\frac{1}{\sqrt{3}} \begin{pmatrix}
 1 &   1   & 1  \\
1 & \omega^{2}   & \omega \\
1  &  \omega & \omega^{2}  \\
\end{pmatrix}.\left(\begin{array}{ccc} 1 & 0 & 0 \\ 0 & \cos\theta_{\nu} & \sin\theta_{\nu} e^{i\delta_{\nu}} \\ 0 & -\sin\theta_{\nu} e^{-i\delta_{\nu}} & \cos\theta_{\nu} \\ \end{array} \right)\,.
\end{eqnarray}
One sees from Eq.~\eqref{eq:Umix} that one column of lepton mixing matrix is fixed to be $\frac{1}{\sqrt{3}}(1,1,1)^{T}$.
Since in this model, we have no prediction for the ordering of neutrino mass eigenvalues, we are allowed to permute the columns in $U_{}$.
This way one sees that the lepton mixing matrix has the so-called trimaximal $\text{TM}2$ form~\cite{Albright:2008rp,Albright:2010ap,He:2011gb},
\begin{equation}
\label{eq:U_{PMNS}}
U_{}'=U_{}.P_{12}=\frac{1}{\sqrt{3}}\left( \begin{array}{ccc} \cos\theta_{\nu}-\sin\theta_{\nu}e^{-i\delta_{\nu}} & 1 & \cos\theta_{\nu}+\sin\theta_{\nu}e^{i\delta_{\nu}}\\
\omega^{2}\cos\theta_{\nu}-\omega \sin\theta_{\nu}e^{-i\delta_{\nu}} & 1 & \omega \cos\theta_{\nu}+\omega^{2}\sin\theta_{\nu}e^{i\delta_{\nu}}\\
\omega \cos\theta_{\nu}-\omega^{2}\sin\theta_{\nu}e^{-i\delta_{\nu}} & 1 & \omega^{2}\cos\theta_{\nu}+\omega \sin\theta_{\nu}e^{i\delta_{\nu}} \\ \end{array} \right)\,,
\end{equation}
where
\begin{equation}
  P_{12}=\begin{pmatrix} 0 & 1 & 0 \\ 1 & 0 & 0 \\ 0 & 0 & 1 \end{pmatrix}\,.
\end{equation}
We find that the lepton mixing matrix $U'$ in Eq.~\eqref{eq:U_{PMNS}} fulfills the following identities,
\begin{equation}
U'(\theta_{\nu}+\pi, \delta_{\nu})=-U'(\theta_{\nu}, \delta_{\nu}),~~~U'(\theta_{\nu}, \delta_{\nu}+\pi)=U'(-\theta_{\nu}, \delta_{\nu})\,.
\end{equation}
Consequently the parameters $\theta_{\nu}$ and $\delta_{\nu}$ can be restricted to the regions $0\leq\theta_{\nu}\leq\pi$ and $0\leq\delta_{\nu}\leq\pi$ without loss of generality.

\subsection{Oscillation predictions}
\label{sec:oscill-pred}

As already anticipated at the end of Sec.~\ref{sec:neutrino_mass}, we expect two predicions for the oscillation parameters in our scenario.
Indeed, from the lepton mixing matrix obtained in Eq.~\eqref{eq:U_{PMNS}}, one can easily extract the following results for the neutrino mixing angles as well as the leptonic Jarlskog invariant,
\begin{eqnarray}
\nonumber&&\sin^{2}\theta_{13}=\frac{1+\sin2\theta_{\nu}\cos\delta_{\nu}}{3}\,,\\
\nonumber&&\sin^{2}\theta_{12}=\frac{1}{2-\sin2\theta_{\nu}\cos\delta_{\nu}}\,,\\
\nonumber&&\sin^{2}\theta_{23}=\frac{1}{2}-\frac{\sqrt{3}\sin2\theta_{\nu}\sin\delta_{\nu}}{4-2\sin2\theta_{\nu}\cos\delta_{\nu}}\,,\\
&&J_{CP}=\frac{\cos2\theta_{\nu}}{6\sqrt{3}}\,.
\end{eqnarray}
One sees that the three neutrino mixing angles as well as the Dirac CP violation phase are all expressed in terms of just two parameters, $\theta_{\nu}$ and $\delta_{\nu}$.
As a result, there are two predicted relations amongst the neutrino mixing angles and the Dirac CP violation phase, that can be expressed analytically as
\begin{equation}
\label{eq:osc-predictions}
\cos^{2}\theta_{13}\sin^{2}\theta_{12}=\frac{1}{3}\,,\quad \cos\delta_{CP}=\frac{2(3\cos ^{2}\theta_{12}\cos^{2}\theta_{23}+3\sin^{2}\theta_{12}\sin^{2}\theta_{13}\sin^{2}\theta_{23}-1)}{3\sin 2\theta_{23}\sin2\theta_{12}\sin\theta_{13}}\,.
\end{equation}
Note that the $3\sigma$ region of the solar mixing angle obtained from the latest neutrino oscillation global study is $0.271\leq\sin^2\theta_{12}\leq 0.370$~\cite{deSalas:2020pgw}.
Using the $3\sigma$ range $2.015\times10^{-2}\leq\sin^2\theta_{13}\leq2.417\times10^{-2}$ for NO and  $2.039\times10^{-2}\leq\sin^2\theta_{13}\leq2.441\times10^{-2}$ for IO obtained for the very precisely measured reactor angle $\theta_{13}$~\cite{deSalas:2020pgw}, we find that only a narrow range is consistent in our model,
\begin{eqnarray}
\nonumber&& \text{NO:} ~~ 0.3402 \leq \sin^2\theta_{12}\leq 0.3416\,,\\
\label{eq:solar}&&\text{IO:}~~  0.3403 \leq \sin^2\theta_{12}\leq 0.3417\,,
\end{eqnarray}
a prediction which should be tested in forthcoming neutrino oscillation experiments.
Notice that, since there is no free mixing parameter associated to the diagonalization of the charged leptons, we have this tight prediction.
\begin{figure}[h!]
\centering
\begin{tabular}{cc}
\includegraphics[width=0.49\linewidth]{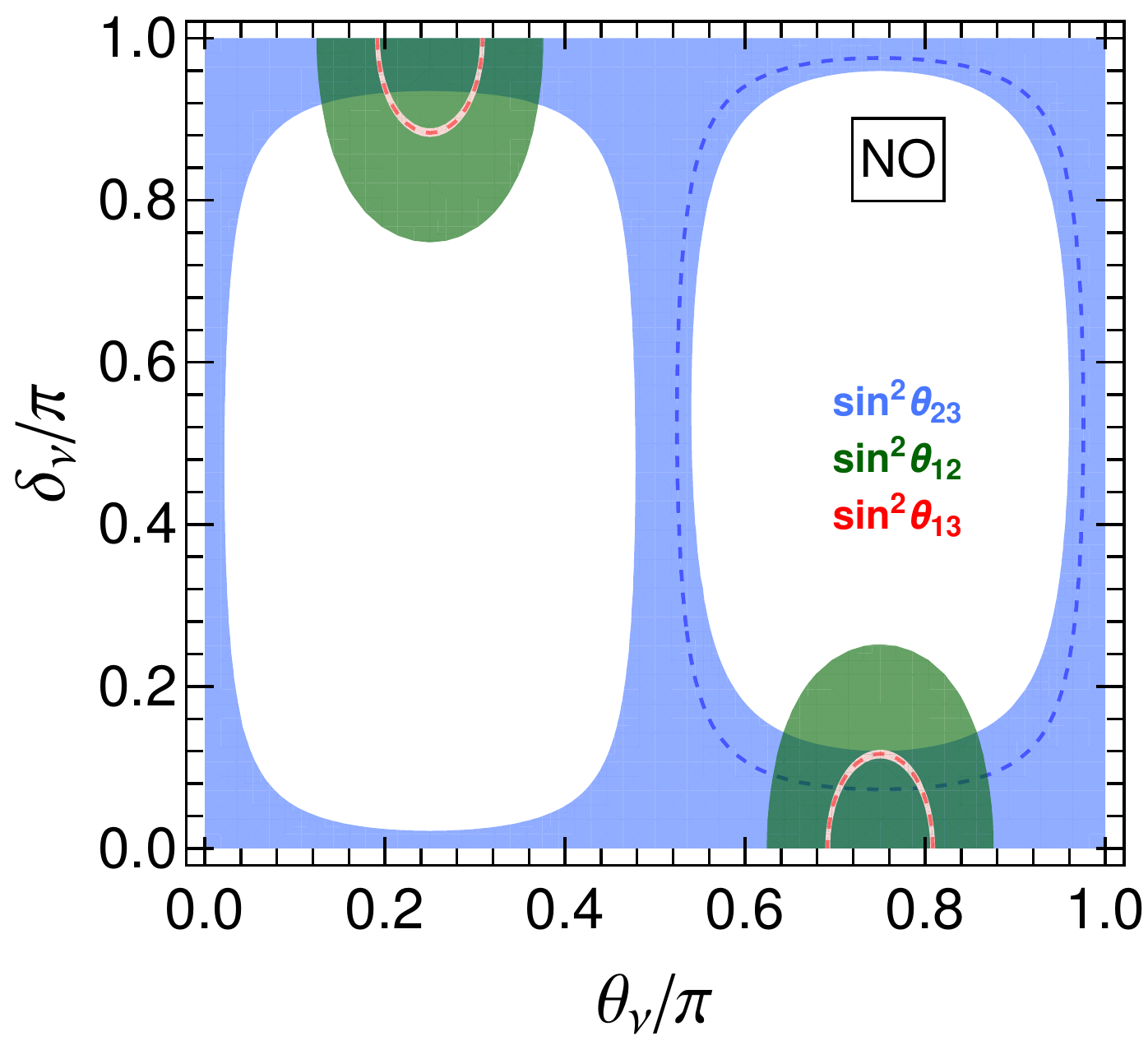}
\includegraphics[width=0.49\linewidth]{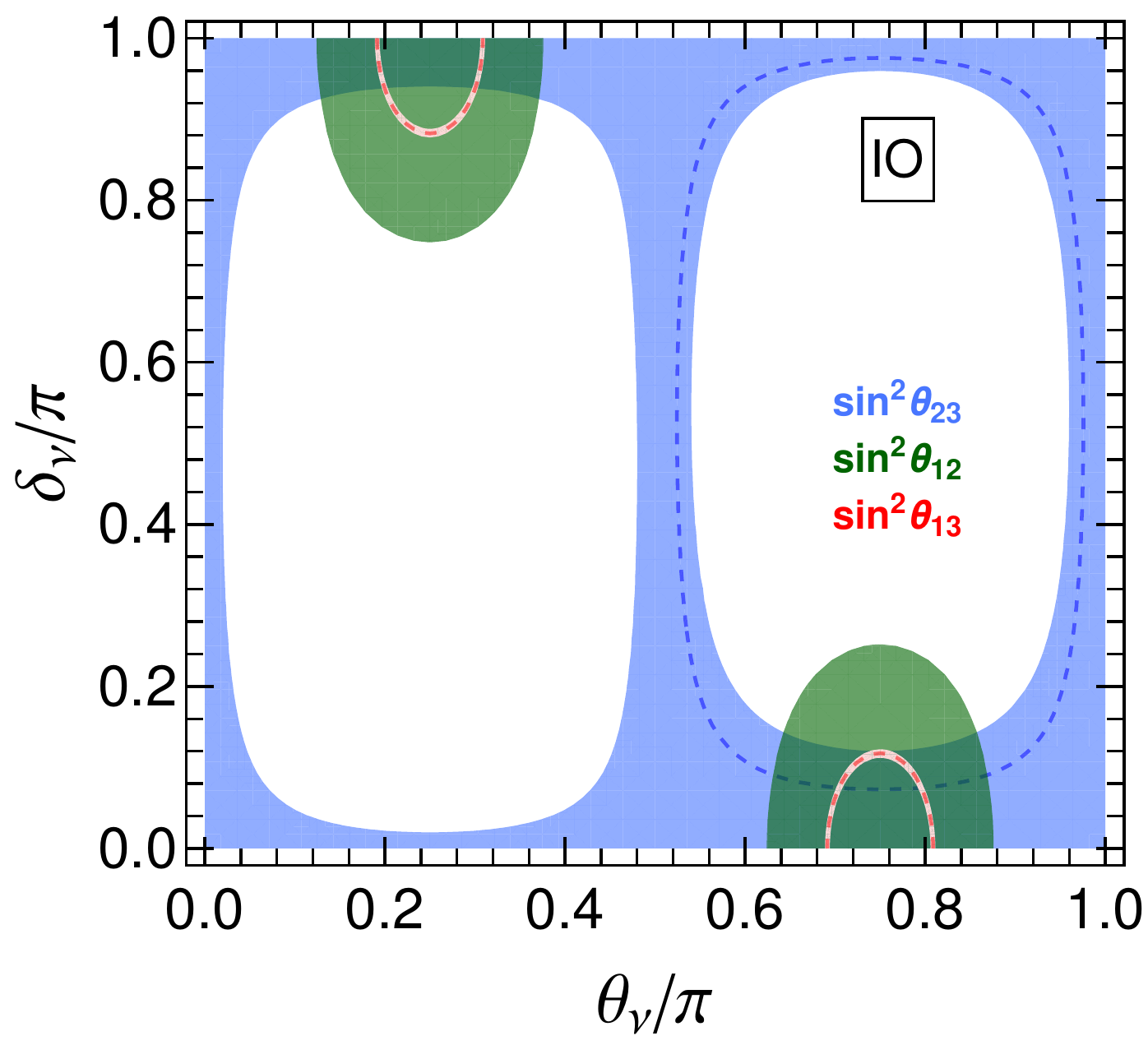}
\end{tabular}
\caption{\label{fig:osc-predictions-I} Contour plots of $\sin^2\theta_{12}$, $\sin^2\theta_{13}$, and $\sin^{2}\theta_{23}$ in the $\theta_{\nu}-\delta_{\nu}$ plane. The red, green and blue areas denote the $3\sigma$ regions of $\sin^{2}\theta_{13}$, $\sin^{2}\theta_{12}$ and $\sin^{2}\theta_{23}$ respectively, and the dashed lines refer to their best fit values taken from \cite{deSalas:2020pgw}.}
\end{figure}
\begin{figure}[t!]
\centering
\begin{tabular}{c}
\includegraphics[width=0.52\linewidth]{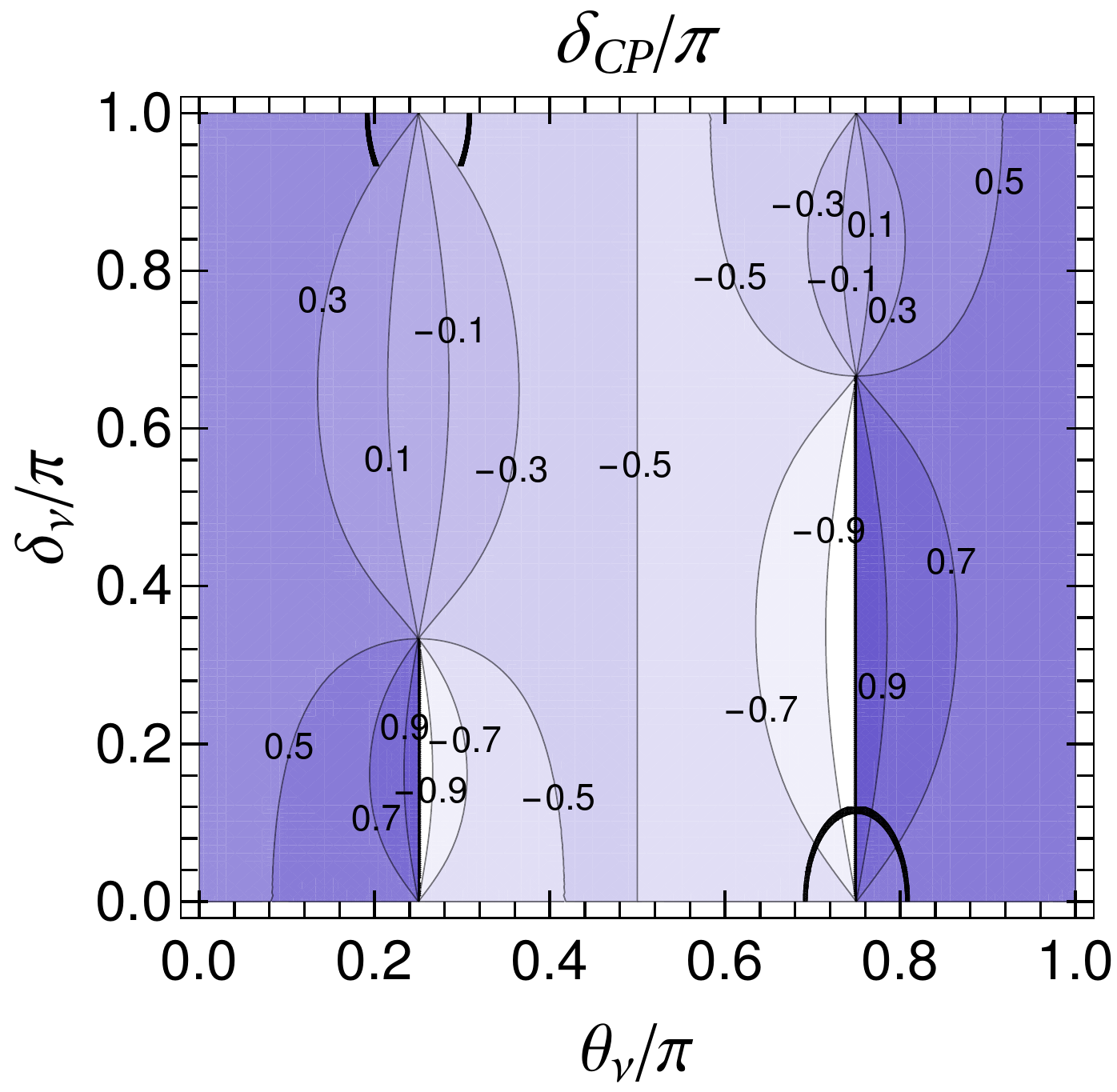}
\end{tabular}
\caption{\label{fig:osc-predictions-II} Contour plots of $\delta_{CP}$ in the $\theta_{\nu}-\delta_{\nu}$ plane. The black areas correspond to the  $3\sigma$ allowed regions of lepton
  mixing angles~\cite{deSalas:2020pgw}. }
 \end{figure}
 \begin{figure}[h!]
\centering
\begin{tabular}{cc}
\includegraphics[width=0.48\linewidth]{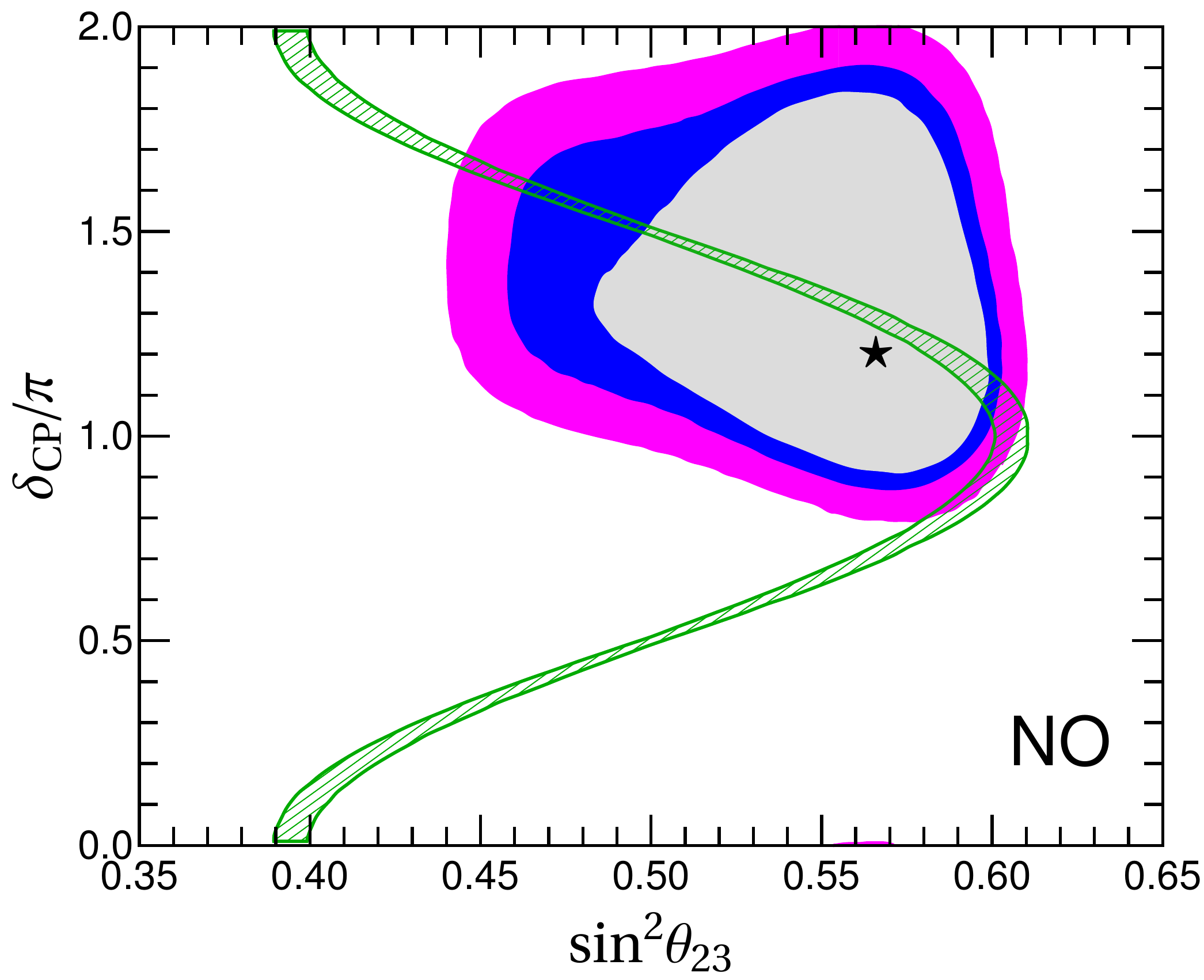}
\includegraphics[width=0.48\linewidth]{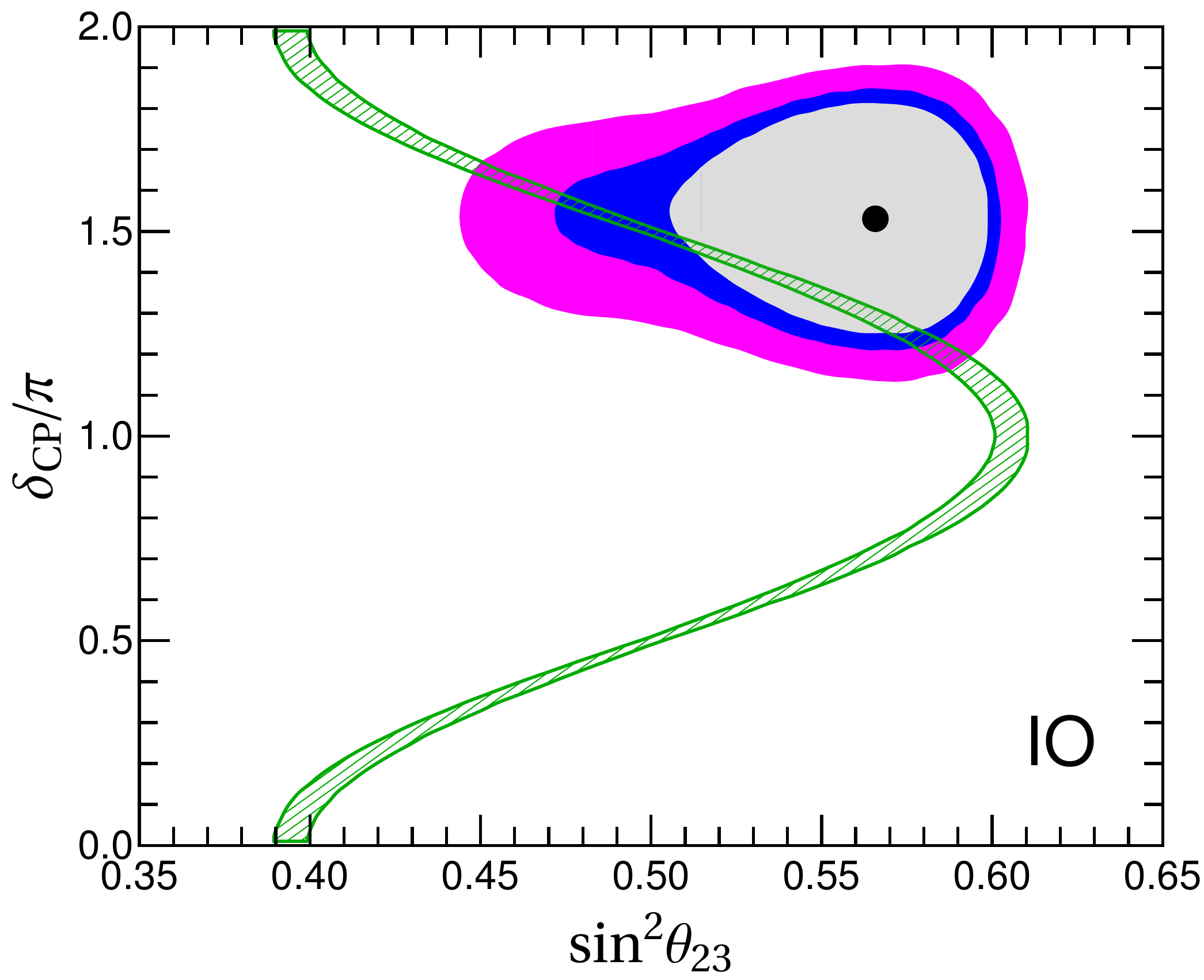}\\
\includegraphics[width=0.48\linewidth]{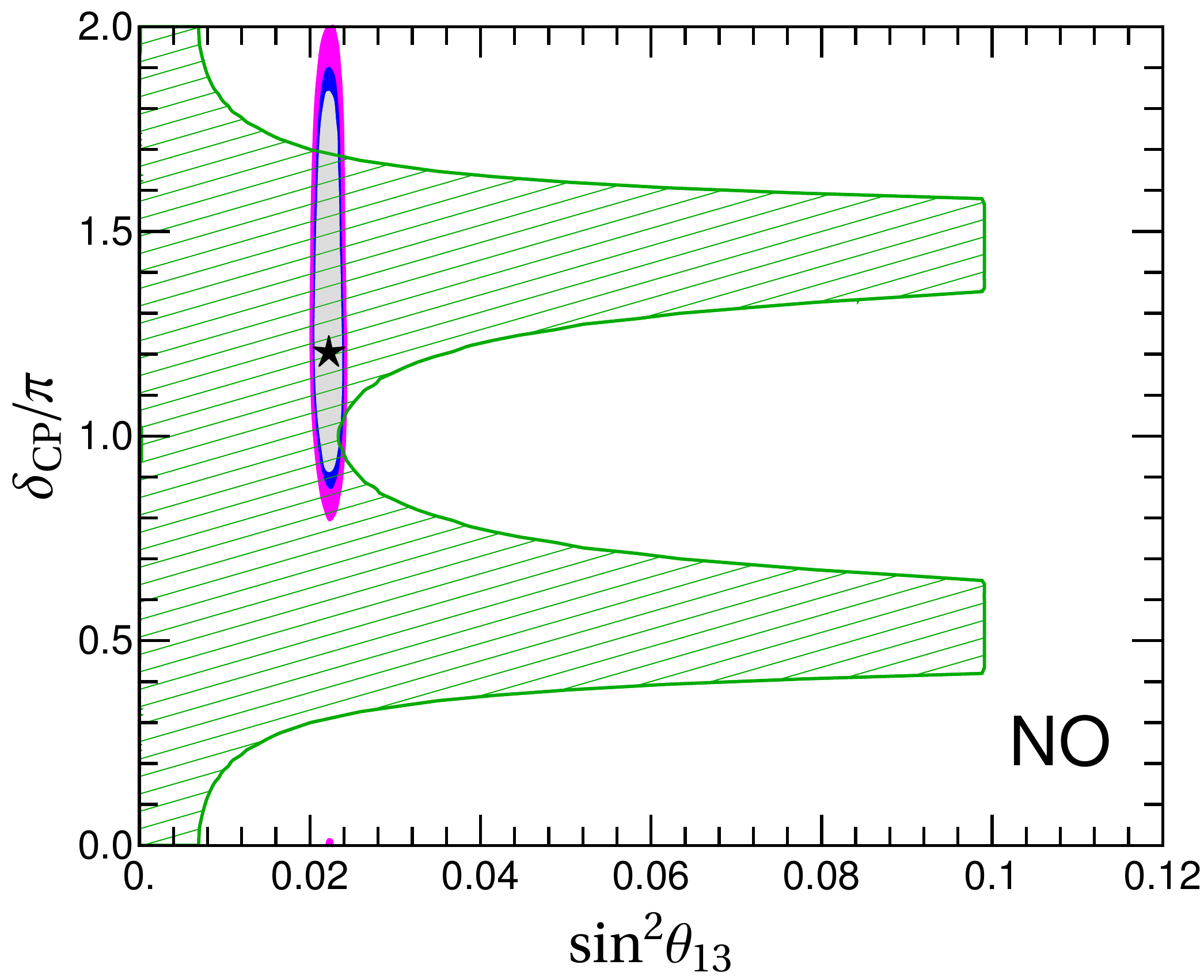}
\includegraphics[width=0.48\linewidth]{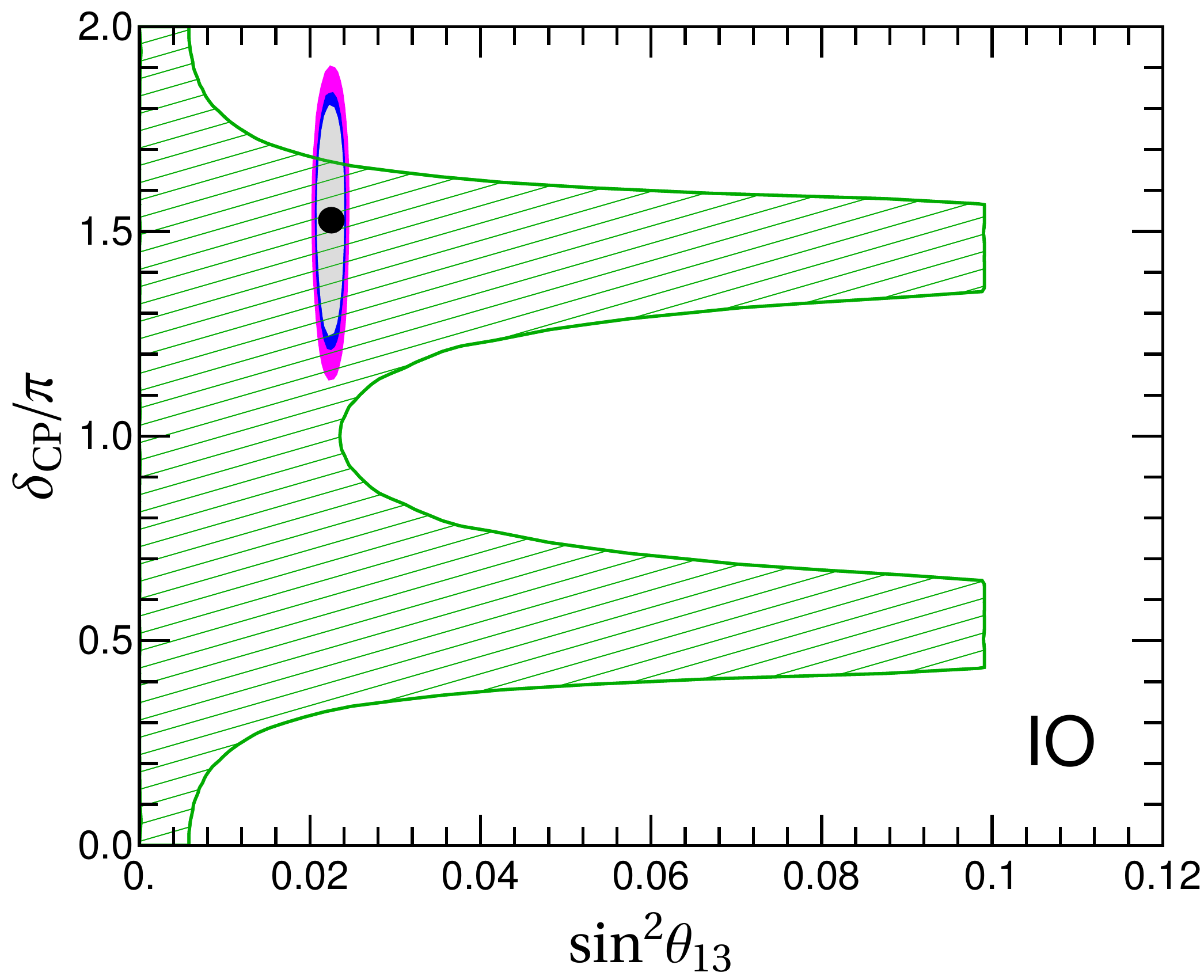}\\
\end{tabular}
\caption{\label{fig:osc-predictions-III} 
  The predicted correlations between $\delta_{CP}$ and $\sin^{2}\theta_{23}$ are shown in the hatched bands in the top panels, for both neutrino orderings, as indicated.
The undisplayed $\sin^{2}\theta_{13}$ and $\sin^{2}\theta_{12}$ are required to lie within their $3\sigma$ regions from the current oscillation global fit~\cite{deSalas:2020pgw}.
Similarly, in displaying $\delta_{CP}$ versus $\sin^{2}\theta_{13}$ we required $\sin^{2}\theta_{12}$ and $\sin^{2}\theta_{23}$ to satisfy their $3\sigma$ constraints.
The $1\sigma$, $2\sigma$ and  $3\sigma$ regions from the current neutrino oscillation global fit are indicated by the shaded areas~\cite{deSalas:2020pgw}. }
\end{figure}

The predicted values of the mixing parameters $\sin^{2}\theta_{ij}$ and $\delta_{CP}$ are determined by scanning the free parameters $\theta_{\nu}$ and $\delta_{\nu}$ from $0$ to $\pi$. The measurement of the oscillation parameters $\sin^2\theta_{12}$, $\sin^2\theta_{13}$, $\sin^2\theta_{23}$ and the Dirac CP violation phase $\delta_{CP}$ place restrictions
 on the plane defined by the model parameters $\theta_{\nu}-\delta_{\nu}$, shown in Figs.~\ref{fig:osc-predictions-I} and \ref{fig:osc-predictions-II}. 
The shaded regions in Fig.~\ref{fig:osc-predictions-I} followd from the individual measurements of the three mixing angles, according to the global oscillation analysis in Ref.~\cite{deSalas:2020pgw}.
Fig.~\ref{fig:osc-predictions-II} shows the corresponding contour plots for $\delta_{CP}$ in the $\theta_{\nu}-\delta_{\nu}$ plane.
The black bands denote the regions in which all three lepton mixing angles lie in the experimentally allowed $3\sigma$ ranges~\cite{deSalas:2020pgw}.

The model predictions for the two most poorly determined oscillation parameters  $\sin^2\theta_{23}$ and $\delta_{CP}$ are shown in Fig.~\ref{fig:osc-predictions-III}.
  The top panels illustrate the very tight correlations between  $\sin^2\theta_{23}$ and $\delta_{CP}$.
  The bottom display $\delta_{CP}$ versus $\sin^{2}\theta_{13}$, the star and dot represent the global best fit points for NO and IO, respectively.
  One sees that the allowed values of $\delta_{CP}$ are ``cut-from-above'', covering a narrower range than that obtained in generic global fit determinations.

\subsection{Neutrinoless double beta decay }
\label{sec:neutr-double-beta}

\begin{figure}[h!]
\centering
\begin{tabular}{c}
\includegraphics[width=0.6\linewidth]{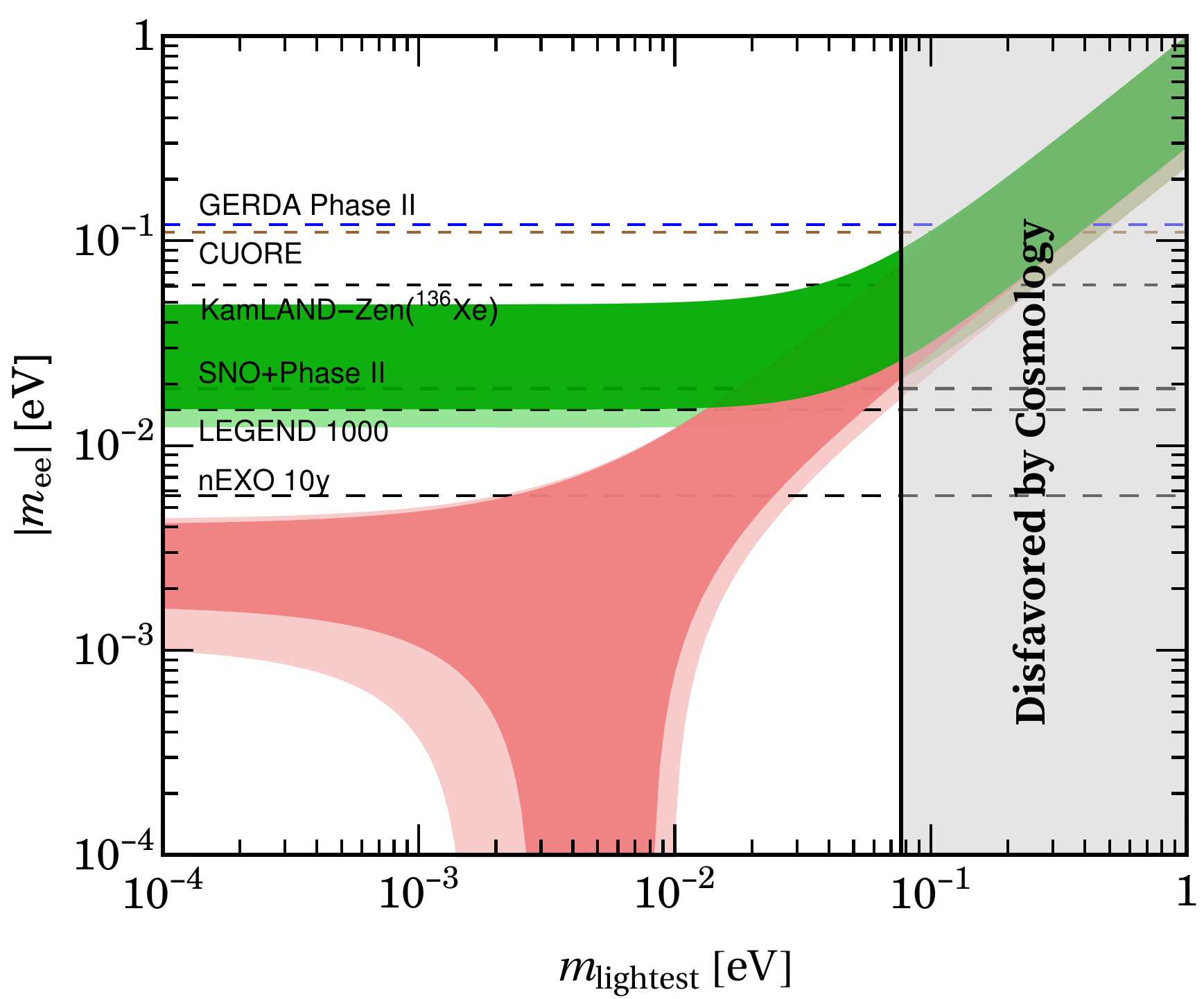}
\end{tabular}
\caption{\label{fig:znbb} 
Expected mass parameter characterizing the $0\nu\beta\beta$ amplitude. The values of the neutrino oscillation parameters are taken from \cite{deSalas:2020pgw}.
The red and blue regions are for IO and NO respectively.
The vertical grey band is excluded by the cosmological limit $\Sigma_{i}m_{i}<0.120\text{eV}$ at $95\%$ CL obtained by the Planck collaboration~\cite{Aghanim:2018eyx}.}
\end{figure}

Given the oscillation results one can forecast the expected values for the mass parameter $|m_{ee}|$ characterizing the amplitude for neutrinoless double beta decay ($0\nu\beta\beta$).
The general expectations cover two regions depending on whether neutrinos mass eigenvalues follow a normal or inverted ordering (NO or IO).
In Fig.~\ref{fig:znbb} we plot the regions expected within our scenario.
As a result of our predictions, Eq.~(\ref{eq:osc-predictions}), they are narrower than expected generically.
Despite this fact, for the NO case, preferred by oscillations, the amplitude can vanish both in the general case as well as within our model.
In contrast, the lower bound (in the amplitude) expected for the IO case lies over 20\% higher in our model than in the generic case.
This would make \znbb detectable at LEGEND~\cite{Abgrall:2017syy} and nEXO~\cite{Albert:2017hjq}.

\section{Final discussion}
\label{sec:final-discussion}

We have proposed a scotogenic theory in which an $A_4$ family symmetry is incorporated.
Our scheme provides a natural way to derive a trimaximal neutrino mixing pattern from first principles.
Trimaximal neutrino mixing had been derived from earlier theories of lepton mixing.
For example, the TM1 pattern was derived in~\cite{Ding:2013hpa}.
More recently, TM1 has been recently derived from constructions involving modular symmetries~\cite{King:2019vhv} or warped extra dimensions~\cite{Chen:2020udk}.
Concerning the TM2 pattern obtained in the present paper, it has been obtained within the warped scenario in Ref.~\cite{Chen:2015jta}.
However, our present 4-dimensional model is substantially simpler and provides also an explanation for cosmological dark matter.
The particle responsible for the latter is a weakly interacting massive particle that mediates neutrino mass generation, Fig.~\ref{fig:loop}.
The dark matter phenomenology has been studied in the literature for the simpler, flavour-less scotogenic model.
In fact, if the lightest dark-sector particle is a scalar, the dark matter analysis is equivalent to the one recently explored in~\cite{Avila:2019hhv}.
Last, but not least, we stress the two very simple, but important, neutrino predictions in Eq.~(\ref{eq:osc-predictions}).
These have been thoroughly discussed in this paper.
They imply a very narrow range for the solar mixing angle, Eq.~(\ref{eq:solar}), and sharp correlations between  $\sin^2\theta_{23}$ and $\delta_{CP}$, as shown in Fig.~\ref{fig:osc-predictions-III}, leading to a restricted consistency range for $\delta_{CP}$.
In addition, we have a somewhat tighter lower bound on the mass parameter characterizing the \znbb amplitude.
These predictions should be testable at future oscillation experiments as well as $0\nu\beta\beta$ searches.\\[-.2cm]

Our model may be extended to include quarks. In the minimal way, without new scalars, we could assign the left-handed quarks to an $A_4$ triplet,
  the right-handed up- and down-quarks as singlets $\mathbf{1}$, $\mathbf{1'}$ and $\mathbf{1''}$ respectively.
  The up- and down-quark mass terms would be of the same form as the charged lepton Yukawa couplings in Eq.~\eqref{eq:charged-lepton-Yukawa}.
  As a result, the CKM mixing matrix would be the unit matrix, a good leading order approximation.
  In order to accommodate the measured CKM pattern, one could include other scalar fields beyond $\phi$, with non-zero vacuum expectation values.

\begin{acknowledgments}

  This work is supported by the Spanish grants SEV-2014-0398 and FPA2017-85216-P (AEI/FEDER, UE), PROMETEO/2018/165 (Generalitat Valenciana),
  Funda\c{c}{\~a}o para a Ci{\^e}ncia e a Tecnologia (FCT, Portugal) under project CERN/FIS-PAR/0004/2019, and the Spanish Red Consolider MultiDark FPA2017-90566-REDC
  and the National Natural Science Foundation of China under Grant Nos 11975224,  11835013 and 1194730.

\end{acknowledgments}


\providecommand{\href}[2]{#2}\begingroup\raggedright\endgroup

\end{document}